\documentclass[showpacs,aps,prl,twocolumn,superscriptaddress]{revtex4}
\usepackage{graphicx} 
\usepackage{dcolumn}
\usepackage{bm}
\usepackage{amssymb,amsmath}
\usepackage{epstopdf}
\usepackage[T1]{fontenc}
\usepackage[latin9]{inputenc}
\usepackage{color}
\usepackage{float}
\usepackage{esint}
\usepackage{graphics}
\usepackage{bm}
\makeatletter
\@ifundefined{textcolor}{}
{%
 \definecolor{BLACK}{gray}{0}
 \definecolor{WHITE}{gray}{1}
 \definecolor{RED}{rgb}{1,0,0}
 \definecolor{GREEN}{rgb}{0,1,0}
 \definecolor{BLUE}{rgb}{0,0,1}
 \definecolor{CYAN}{cmyk}{1,0,0,0}
 \definecolor{MAGENTA}{cmyk}{0,1,0,0}
 \definecolor{YELLOW}{cmyk}{0,0,1,0}
}

\begin{document}
\title{Direct Observation of Cross-Polarized Excitons in \\Aligned Single-Chirality Single-Wall Carbon Nanotubes 
}
\normalsize

\author{Fumiya Katsutani}
\affiliation{Department of Electrical and Computer Engineering, Rice University, Houston, Texas 77005, USA}

\author{Weilu Gao}
\affiliation{Department of Electrical and Computer Engineering, Rice University, Houston, Texas 77005, USA}

\author{Xinwei Li}
\affiliation{Department of Electrical and Computer Engineering, Rice University, Houston, Texas 77005, USA}

\author{Yota Ichinose}
\affiliation{Department of Physics, Faculty of Science and Engineering, Tokyo Metropolitan University, Hachioji, Tokyo 192-0397, Japan}

\author{Yohei Yomogida}
\affiliation{Department of Physics, Faculty of Science and Engineering, Tokyo Metropolitan University, Hachioji, Tokyo 192-0397, Japan}


\author{Kazuhiro Yanagi}
\affiliation{Department of Physics, Faculty of Science and Engineering, Tokyo Metropolitan University, Hachioji, Tokyo 192-0397, Japan}

\author{Junichiro Kono}
\email[]{kono@rice.edu}
\affiliation{Department of Electrical and Computer Engineering, Rice University, Houston, Texas 77005, USA}
\affiliation{Department of Physics and Astronomy, Rice University, Houston, Texas 77005, USA}
\affiliation{Department of Materials Science and NanoEngineering, Rice University, Houston, Texas 77005, USA}

\date{\today}

\begin{abstract}
Optical properties of single-wall carbon nanotubes (SWCNTs) for light polarized parallel to the nanotube axis have been extensively studied, whereas their response to light polarized perpendicular to the nanotube axis has not been well explored.  Here, by using a macroscopic film of highly aligned single-chirality (6,5) SWCNTs, we performed a systematic polarization-dependent optical absorption spectroscopy study.  In addition to the commonly observed angular-momentum-conserving interband absorption of parallel-polarized light, which generates $E_{11}$ and $E_{22}$ excitons, we observed a small but unambiguous absorption peak whose intensity is maximum for perpendicular-polarized light.  We attribute this feature to the lowest-energy cross-polarized interband absorption processes that change the angular momentum along the nanotube axis by $\pm 1$, generating $E_{12}$ and $E_{21}$ excitons.  The energy difference between the $E_{12}$ and $E_{21}$ exciton peaks, expected from asymmetry between the conduction and valence bands, was smaller than the observed linewidth.  Unlike previous observations of cross-polarized excitons in polarization-dependent photoluminescence and circular dichroism spectroscopy experiments, our direct observation using absorption spectroscopy allowed us to quantitatively analyze this resonance.  Specifically, we determined the energy and oscillator strength of this resonance to be 1.54 and 0.05, respectively, compared with the values for the $E_{11}$ exciton peak.  These values, in combination with comparison with theoretical calculations, in turn led to an assessment of the environmental effect on the strength of Coulomb interactions in this aligned single-chirality SWCNT film.
\end{abstract}

\pacs{}

\maketitle

\section{Introduction}

\begin{figure}[ht]
	\includegraphics[width=\linewidth]{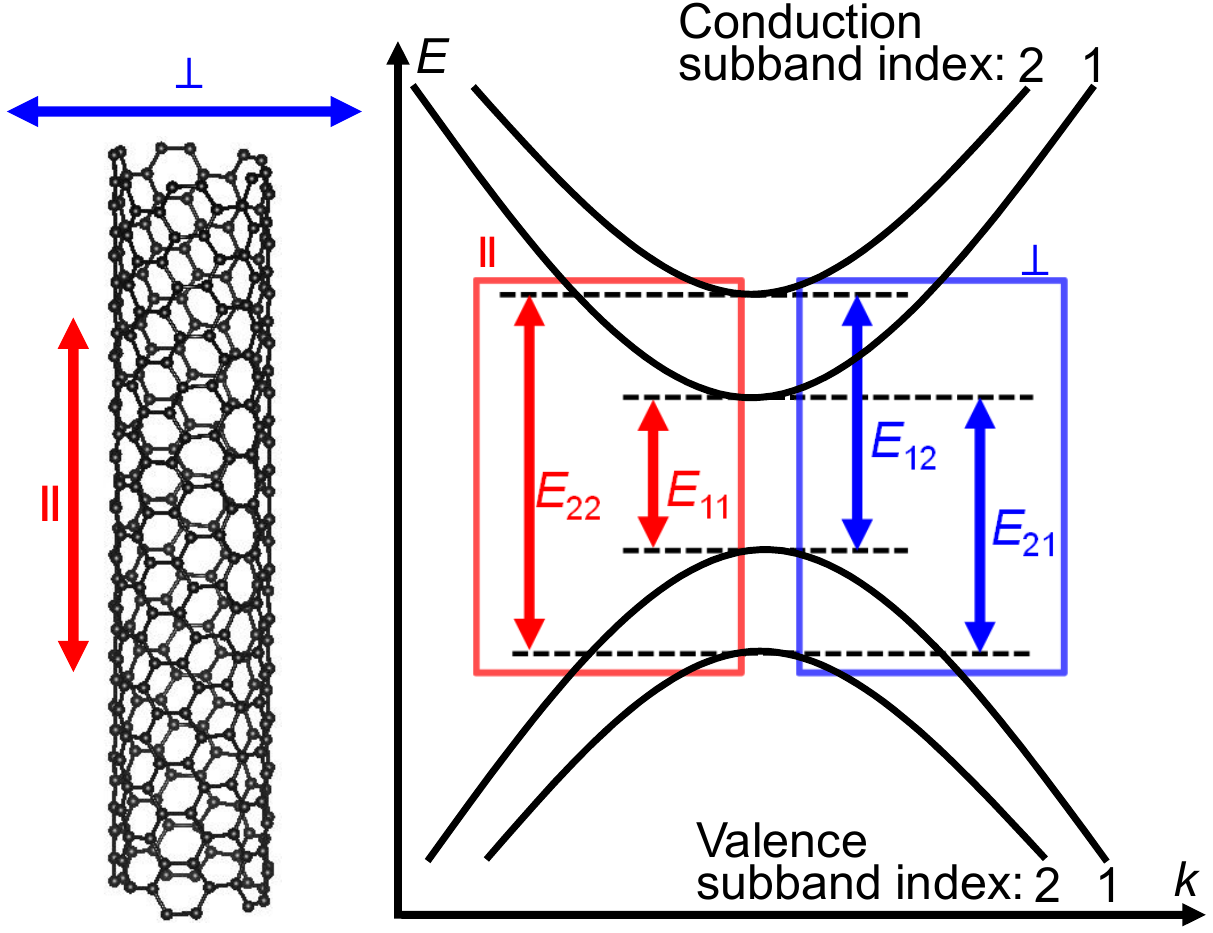}
	\caption{Illustration of the lowest-energy allowed optical interband transitions in a semiconducting SWCNT. The numbers shown for the four subbands, two in the condiuction band and two in the valence band, are their subband indices. $E_{ij}$ ($i = j$) denotes an allowed optical transition for parallel ($\parallel$) polarization, whereas $E_{ij}$ ($i \neq j$) indicates an allowed optical transition for perpendicular ($\perp$) polarization.}
	\label{I_schematic}
\end{figure}

Semiconducting single-wall carbon nanotubes (SWCNTs) possess rich optical properties arising from one-dimensional excitons with extremely large binding energies \cite{Ando97JPSJ,KaneMele03PRL,ChangetAl04PRL,SpataruetAl04PRL,PerebeinosetAl04PRL,ZhaoMazumdar04PRL,WangetAl05Science,MaultzschetAl05PRB2,DukovicetAl05NL}. Although much has been understood about the properties of excitons that are active for parallel-polarized light, excitons excited by perpendicular-polarized light have not been explored experimentally. Such cross-polarized excitons are predicted to exhibit strong many-body effects due to a subtle interplay of quantum confinement and Coulomb interactions \cite{AjikiAndo94Physica,UryuAndo06PRB,UryuAndo07PRB,KilinaetAl08PNAS}.

Figure \ref{I_schematic} schematically shows the lowest-energy allowed interband optical transitions in a semiconducting SWCNT \cite{Ando05JPSJ}. For absorption of light polarized parallel to the nanotube axis, the band index is preserved in an allowed optical transition (the $E_{11}$ and $E_{22}$ transitions). For light polarized perpendicular to the nanotube axis, a transition can occur when the subband index changes by 1 (the $E_{12}$ and $E_{21}$ transitions). As first pointed out by Ajiki and Ando \cite{AjikiAndo94Physica}, the $E_{12}$ and $E_{21}$ absorption peaks are expected to be suppressed  because of the depolarization effect. However, subsequent theoretical studies~\cite{UryuAndo06PRB,UryuAndo07PRB,KilinaetAl08PNAS} taking into account the electron-hole Coulomb interactions indicated that a small absorption peak due to cross-polarized excitons should still appear.

The $E_{12}$/$E_{21}$ transitions were first observed in polarized photoluminescence excitation spectroscopy studies on aqueous suspensions of SWCNTs~\cite{MiyauchietAl06PRB,ChuangetAl08PRB}. By crossing the polarization of the excitation beam with respect to that of the collection beam, $E_{11}$ photoluminescence due to resonant absorption at the $E_{12}$/$E_{21}$ transition was observed. More recently, in circular dichroism (CD) studies \cite{GhoshetAl10NatNano,WeietAl16NC,AoetAl16JACS}, chirality-sorted nanotubes were further separated into enantiomers based on their ``handedness,'' i.e., (6,5) and (5,6) SWCNTs. CD spectra for enantiomer-sorted nanotubes showed peaks due to $E_{12}$ and $E_{21}$ excitons. However, such cross-polarized exciton transitions have never been directly identified in optical absorption spectra. Therefore, quantitative characterization of $E_{12}$/$E_{21}$ excitons has remained elusive.

Here, we report the direct observation of cross-polarized excitons by absorption spectroscopy. Specifically, we investigated the polarization dependence of optical absorption in a macroscopic film of aligned, single-chirality (6,5) SWCNTs. As the angle between the polarization of the incident beam and the nanotube alignment direction was increased from 0$^\circ$ to 90$^\circ$, a peak due to the $E_{12}$/$E_{21}$ excitons appeared and grew in intensity at the expense of the usual parallel-polarized excitons ($E_{11}$ and $E_{22}$). The energy of the $E_{12}$/$E_{21}$ exciton peak was 660\,meV higher than the $E_{11}$ exciton peak and 250\,meV lower than the $E_{22}$ exciton peak. Together with the nematic order parameter of the aligned SWCNT film determined in the same analysis, these polarization-dependent absorption measurements allowed us to determine the oscillator strength of the $E_{12}$/$E_{21}$ peak quantitatively.

\section{SAMPLES AND EXPERIMENTAL METHODS}

\subsection{Preparation of an aligned single-chirality SWCNT film}

\begin{figure}
	\includegraphics[width=\linewidth]{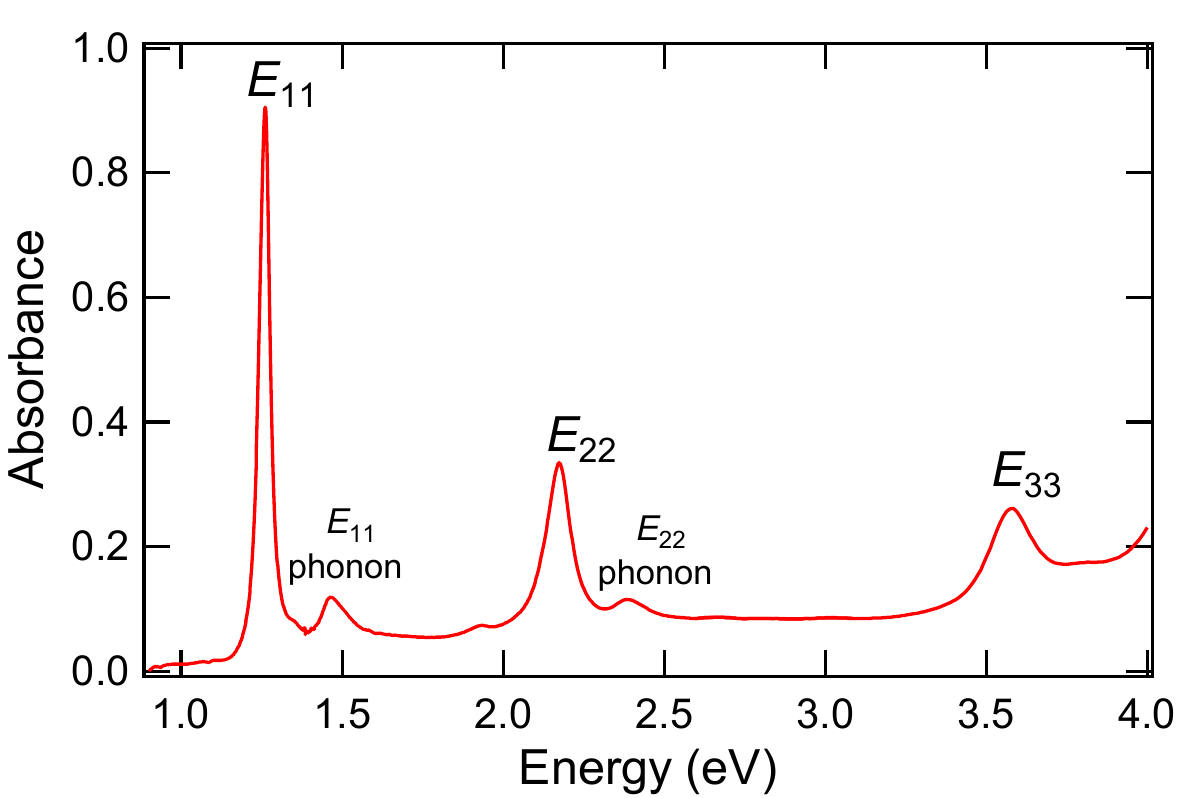}
	\caption{Absorbance spectrum in the near-infrared and visible range for the (6,5)-purified aqueous suspension of SWCNTs with an estimated chirality purity of 99.3\%.  See Appendix~\ref{Appendix:Chirality} for more details on chirality purity determination.}
	\label{II_suspension_spectrum}
\end{figure}

We first prepared an aqueous suspension of extremely pure (6,5) SWCNTs based on pH-controlled gel chromatography~\cite{YomogidaetAl16NC,IchinoseetAl17JPCC}. SWCNTs purchased from Sigma-Aldrich (Signis SG65i) were suspended in an aqueous solution of sodium cholate (SC). After ultracentrifugation, the supernatant was collected as an initial suspension. Sodium dodecyl sulfate (SDS) was added
to the suspension, which was used for a two-stage gel chromatography process. In the first-stage of gel chromatography to separate the semiconducting SWCNTs by a difference in chiral angle, the suspension was loaded onto gel beads (GE Healthcare, Sephacryl S-200 HR) under surfactant environment of 2.0\% SDS and 0.5\% SC, and the nonadsorbed fraction containing (6,5) nanotubes was collected as a filtrate. This filtrate was used for the second-stage process to separate the semiconducting SWCNTs by a difference in diameter and to remove residual metallic SWCNTs. Before separation, the surfactant concentrations of the filtrate were adjusted to 0.5\% SDS and 0.5\% SC.  The pH of the solution strongly influences the adsorption of residual metallic SWCNTs~\cite{IchinoseetAl17JPCC}, and thus, we used pH-adjusted surfactant solutions. The pH-adjusted solutions were loaded on gel beads, and the
adsorbed (6,5) SWCNTs were eluted with a stepwise increase of the concentration of sodium deoxycholate (DOC).

Figure \ref{II_suspension_spectrum} shows an absorbance spectrum for a purified (6,5) suspension in a cuvette with a 10-mm path length. The assigned peaks are $E_{11}$ (1.26\,eV), $E_{11}$ phonon sideband (1.46\,eV), $E_{22}$ (2.17\,eV), $E_{22}$ phonon sideband (2.38\,eV), and $E_{33}$ (3.58\,eV). Small unresolved peaks due to residual metallic nanotubes exist in the range of 2.6--3.1\,eV. We estimate the (6,5) chirality purity of the sample to be 99.3\% from this spectrum.  See Appendix~\ref{Appendix:Chirality} for more details about the method we used for chirality purity determination.

The obtained suspension after surfactant exchange was poured into a 1-inch vacuum filtration system with a 80-nm-pore filter membrane to obtain a wafer-scale film of aligned SWCNTs~\cite{HeetAl16NN}. The prepared suspension contained several surfactants, including SC, sodium dodecylbenzenesulfonate (SDBS), and DOC. In order to have a thick film of highly-aligned (6,5) SWCNTs, we needed to have a mono-surfactant suspension. Therefore, we used ultrafiltration to exchange the mixed surfactants to 0.04\% (wt./vol.)\ DOC. The surfactant concentration was also adjusted to below the critical micelle concentration of DOC through ultrafiltration, which is a necessary condition for the controlled vacuum filtration technique we used to prepare an aligned film~\cite{HeetAl16NN}.  The average length of SWCNTs in the prepared suspension before vacuum filtration was $\sim$200\,nm.

The suspension was poured into a funnel with a polycarbonate filter membrane (Nuclepore track-etched polycarbonate hydrophilic membrane). The pressure underneath the membrane was lowered by a mechanical vacuum pump connected to the side arm of a side-arm flask. The filtration speed was adjusted to a rate of 1--2.5\,mL/hour by controlling the valves in the vacuum line. Near the end of the filtration process, the filtration speed was accelerated to $\sim$10\,mL/hour. In this procedure, the filtration speed was also important to achieve spontaneous alignment~\cite{HeetAl16NN}. The obtained circular film had a diameter of $\sim$20\,mm. The thickness of the film gradually varied from the center ($\sim$10\,nm) to the circumference ($\sim$1\,nm). This film was cut into 4 quadrants. One of them was transferred onto a 1-mm-thick glass substrate by dissolving the filter membrane in chloroform.

\subsection{Polarization-dependent visible--near-infrared absorption spectroscopy}
We performed optical transmission measurements on the prepared SWCNT film using linearly polarized light. Our experimental setup consisted of a tungsten-halogen lamp (Thorlabs, SLS201L), a Glan-Thompson polarizer, and two spectrometers. One of the spectrometers covered a spectral range of 520--1050\,nm, utilizing a monochromator (Horiba/JY, Triax320) equipped with a liquid-nitrogen-cooled CCD camera (Princeton Instruments, Spec-10). The other spectrometer, which covered a spectral range of 1050--1550\,nm, consisted of a monochromator (Princeton Instruments, SP-2150) and a liquid-nitrogen-cooled 1D InGaAs detector array (Princeton Instruments, OMA V InGaAs System). Polarization dependence was achieved through changing the polarization angle of the incident light beam by rotating the polarizer. The light beam was focused down to 30\,$\mu$m in diameter by a 50$\times$ objective lens (Mitutoyo, M Plan NIR 50).

\begin{figure}	
	\includegraphics[width=\linewidth]{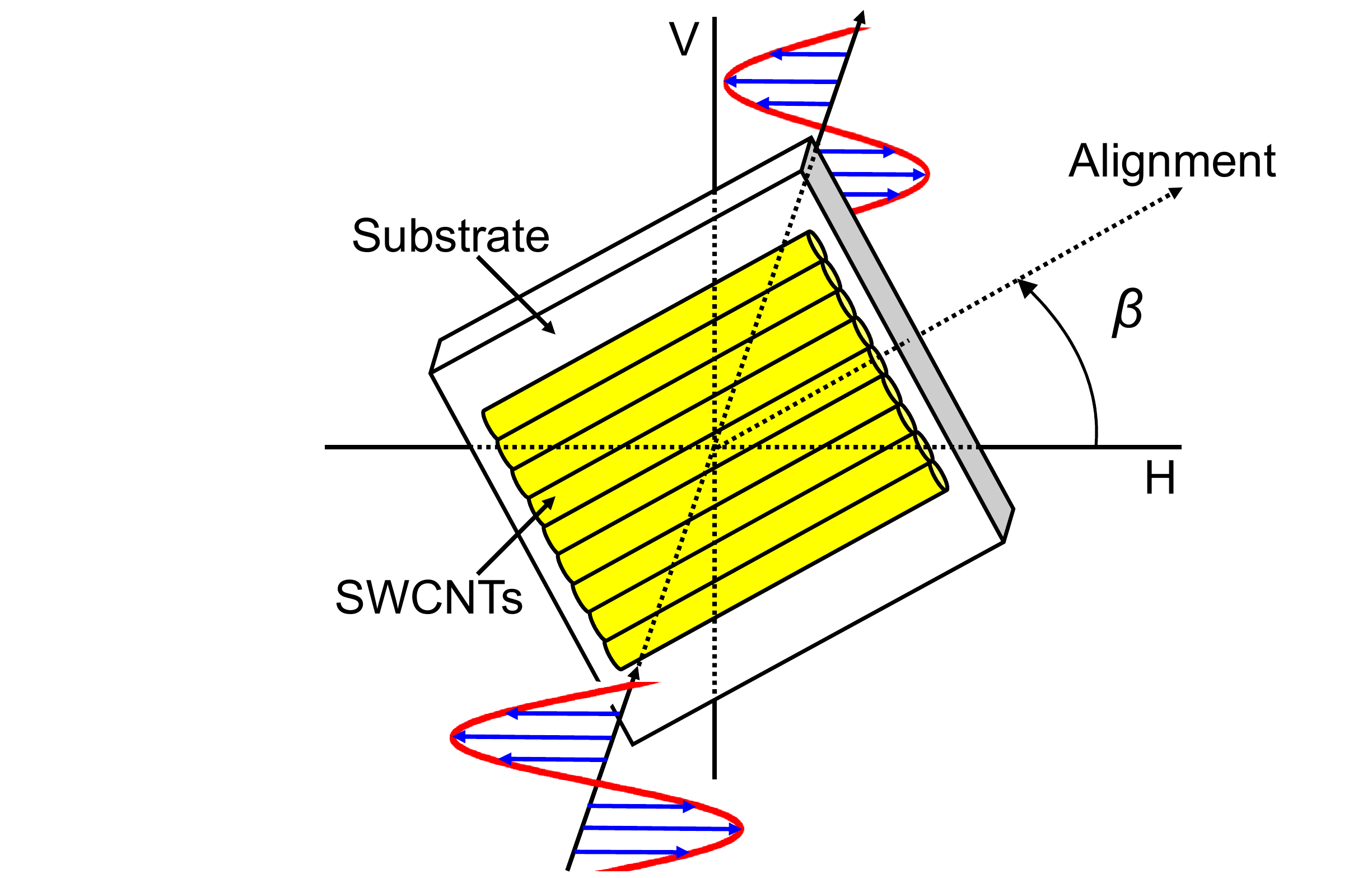}
	\caption{Illustration of the geometry of the polarization-dependent transmission experiments performed on an aligned SWCNT film. The incident beam is linearly polarized along the horizontal axis, and the nanotube alignment direction is rotated from the horizontal axis by angle $\beta$.}
	\label{II_setup}
\end{figure}

A schematic diagram of the experimental geometry is shown in Fig.\,\ref{II_setup}. The incident beam was polarized along the horizontal direction.  The angle between the nanotube alignment direction and the light polarization direction is denoted by $\beta$ throughout this manuscript. Polarization-dependent transmittance ($T$) spectra were taken with a step size of 5 degrees. The measured spot was $\sim$1\,mm away from the center of the film, and the film thickness was $\sim$10\,nm at that spot.  We calculated attenuation spectra through $A = -\ln{(T)}$.

\begin{figure*}
	\includegraphics[width=\linewidth]{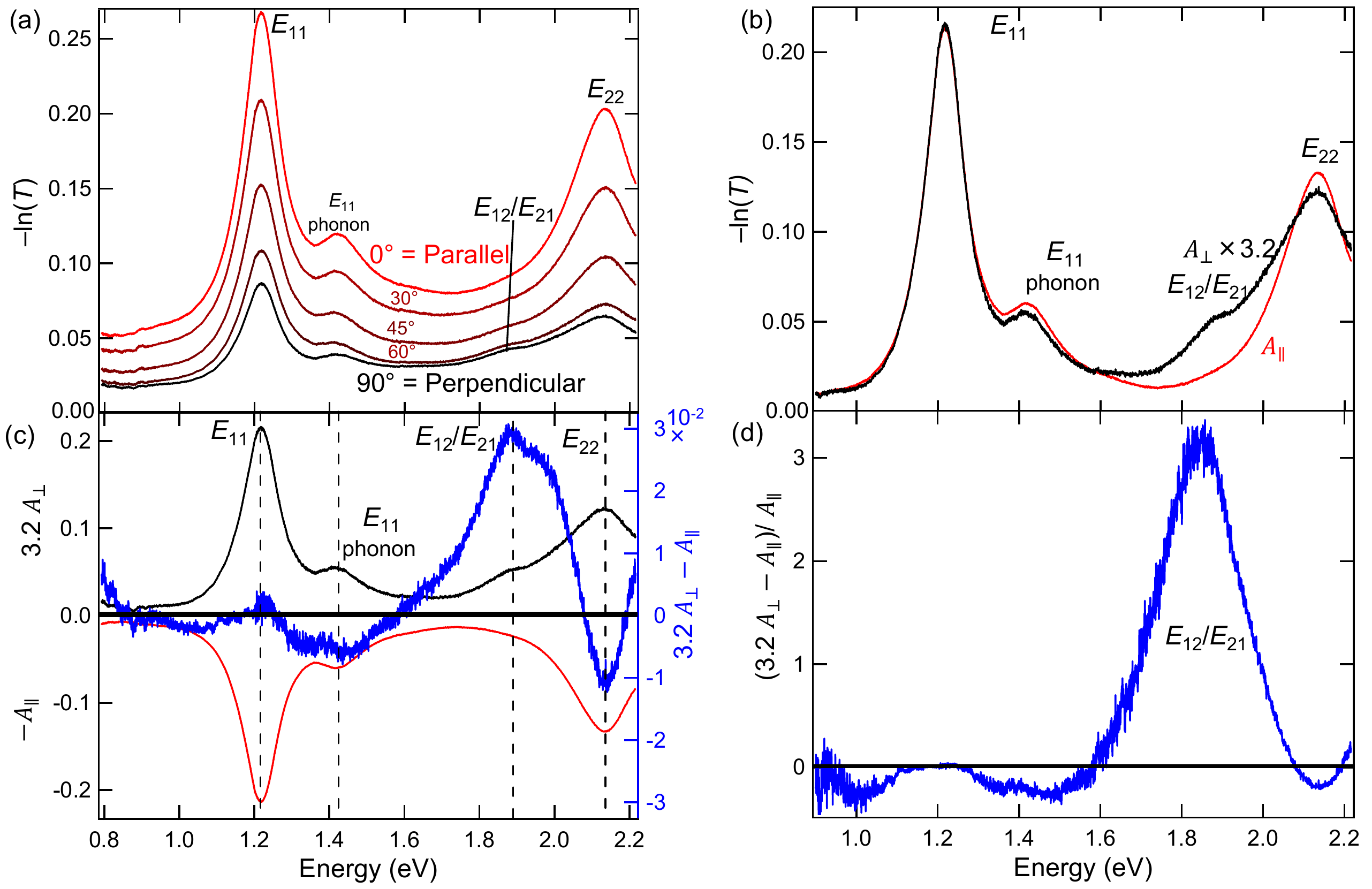}
	\caption{(a)~Polarization-dependent attenuation spectra for the aligned (6,5) SWCNT film for polarization angles ($\beta$) of $0^{\circ}$, $30^{\circ}$, $45^{\circ}$, $60^{\circ}$, and $90^{\circ}$ with respect to the nanotube alignment direction. (b)~Comparison of attenuation spectra for $0^{\circ}$ ($A_\parallel$, black line) and  $90^{\circ}$ ($A_\perp$, red line). $A_\perp$ is multiplied by 3.2. They match except in the spectral region of $E_{12}/E_{21}$.  (c)~Comparison of the $0^{\circ}$ ($A_\parallel$) and $90^{\circ}$ ($A_\perp$) spectra. The blue line indicates $3.2A_\perp - A_\parallel$. (d)~A normalized spectral difference ($3.2A_\perp-A_\parallel$)/$A_\parallel$, which shows a prominent peak due to the $E_{12}/E_{21}$ exciton.}
	\label{III_spectra}
\end{figure*}

\begin{figure*}
	\includegraphics[width=\linewidth]{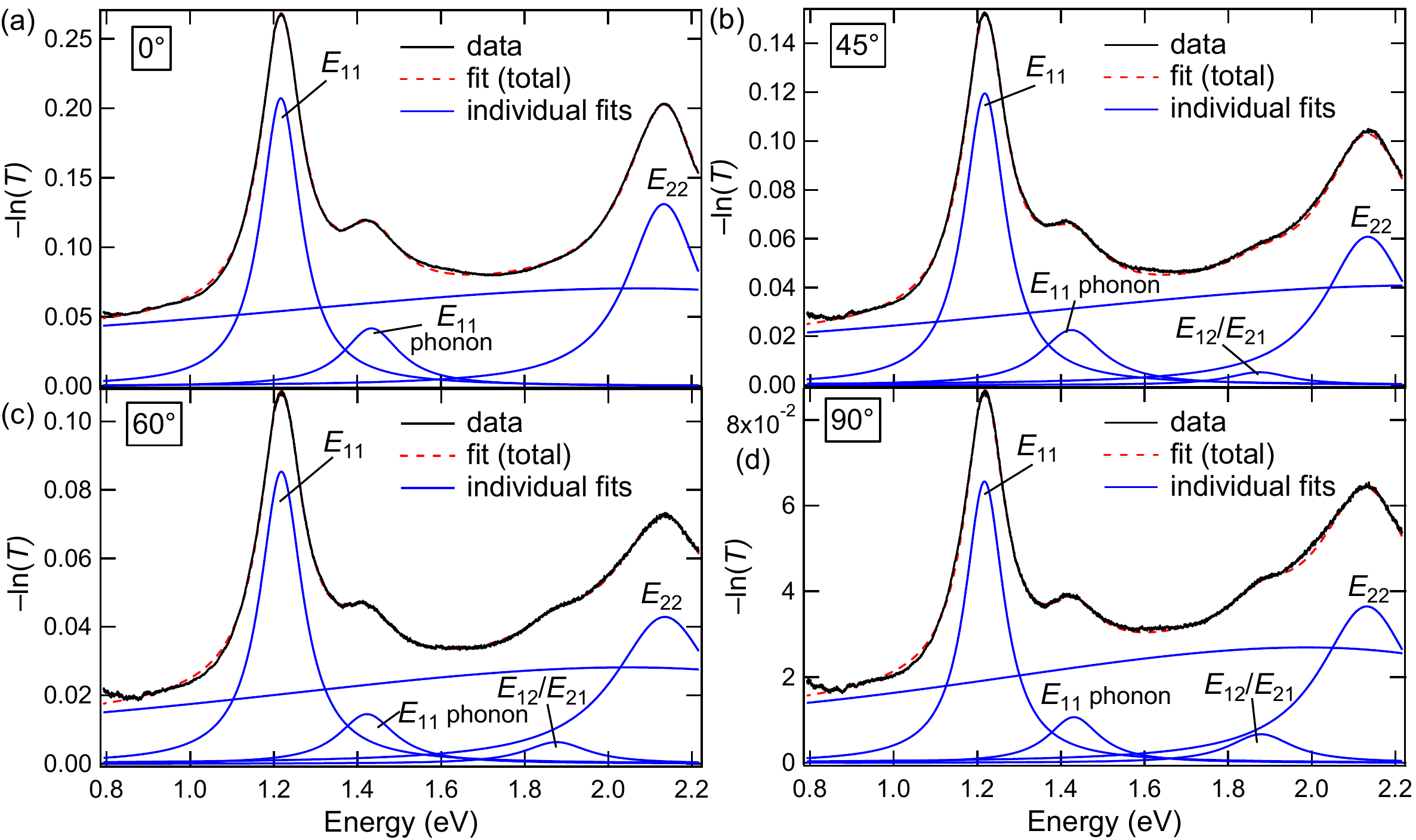}
	\caption{Spectral analysis for the polarization-dependent extinction spectra for the aligned (6,5) SWCNT film using Eq.(\ref{eq_fitting}) as the fit function. The experimental spectra (black), overall fit (red dashed line), and individual components (blue lines) are shown for polarization angles of (a)~0, (b)~45, (c)~60, and (d)~90 degrees.}
	\label{IV_fit}
	\includegraphics[width=\linewidth]{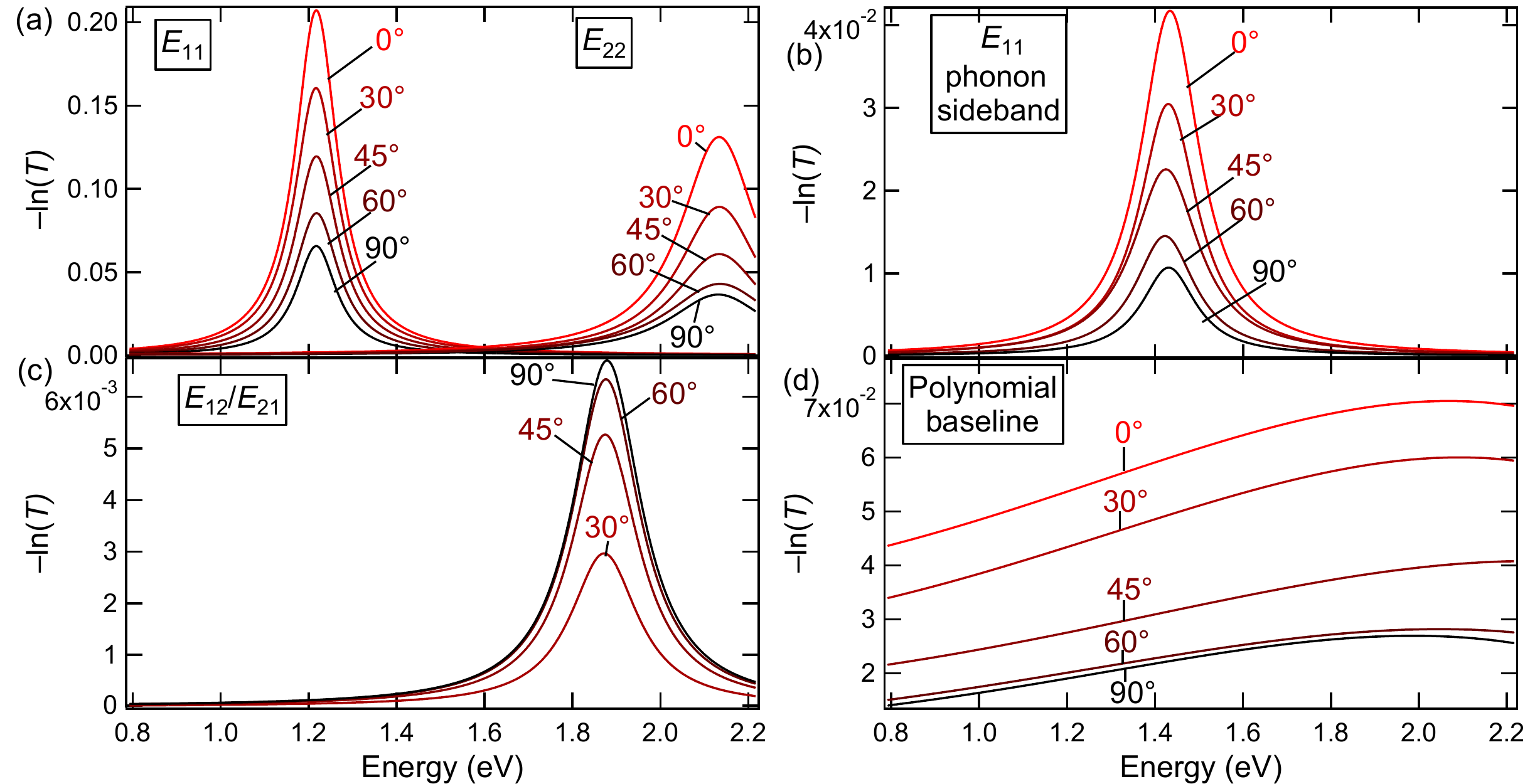}
	\caption{Detailed polarization dependence of the individual spectral components deduced from the fits: (a)~$E_{11}$ and $E_{22}$, (b)~$E_{11}$ phonon sideband, (c)~$E_{12}/E_{21}$, and, (d)~polynomial baseline for polarization angles of 0, 30, 45, 60, 90 degrees.}
	\label{IV_fit_individual}
\end{figure*}

\begin{figure}[tb]
	\includegraphics[width=\linewidth]{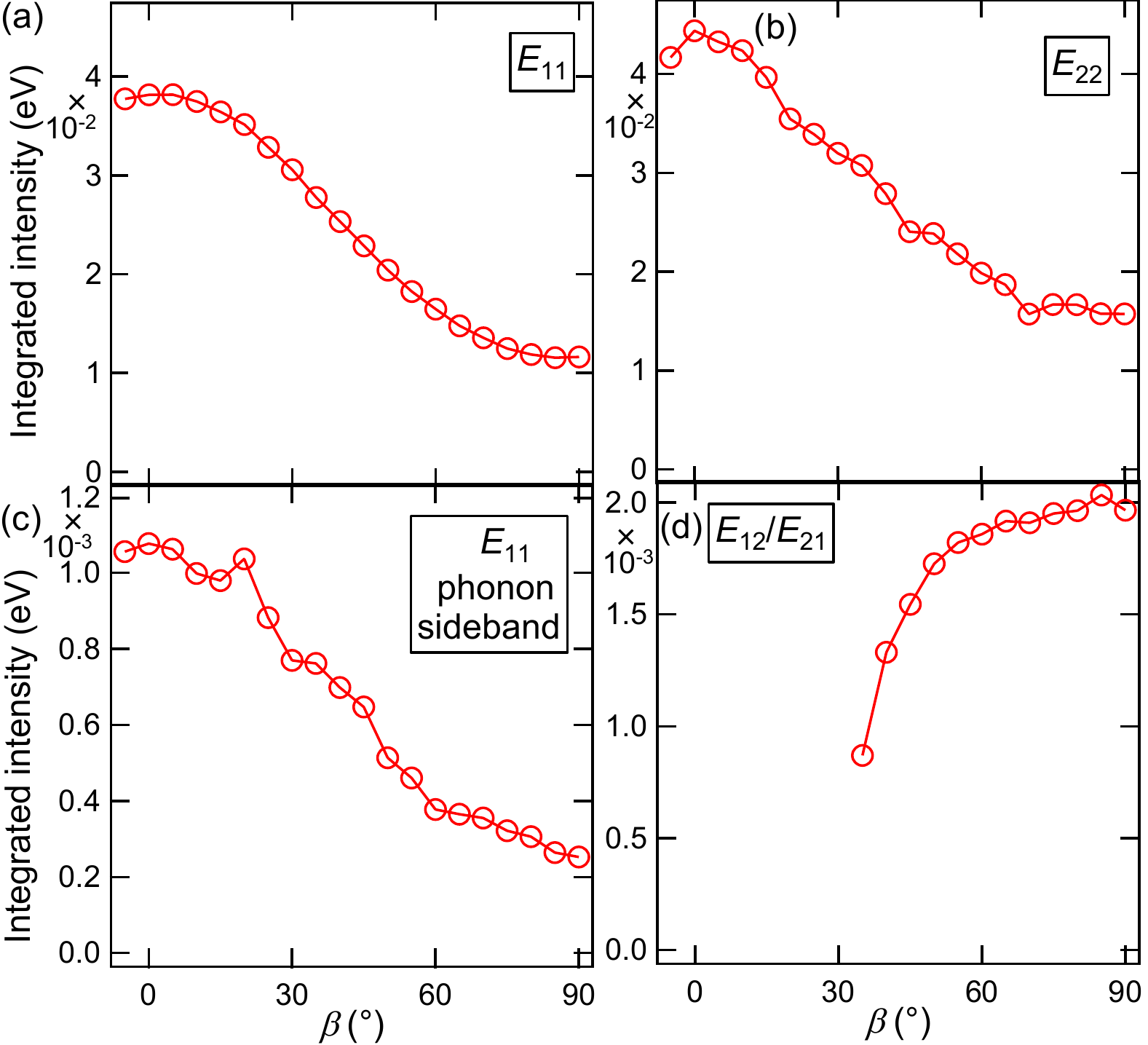}
	\caption{Integrated peak intensity as a function of polarization angle $\beta$ extracted for (a)~$E_{11}$, (b)~$E_{22}$, (c)~$E_{11}$ phonon sideband, and (d)~$E_{12}/E_{21}$.}
	\label{IV_peak_integral}
\end{figure}

\section{EXPERIMENTAL RESULTS}
Figure \ref{III_spectra}(a) displays representative attenuation spectra for polarization angles $\beta$ = 0$^{\circ}$, 30$^{\circ}$, 45$^{\circ}$, 60$^{\circ}$, and 90$^{\circ}$. The spectra are not intentionally offset. The observed peaks at 1.22\,eV and 2.13\,eV are due to the $E_{11}$ and $E_{22}$ exciton transitions, respectively. These peaks are red-shifted compared with the suspension spectrum in Fig.\,\ref{II_suspension_spectrum} by  $\sim$40\,meV . The peak at 1.44\,eV is the phonon sideband of the $E_{11}$ exciton peak. No other peaks are observed due to any residual semiconducting chiralities within this energy range. As the polarization angle $\beta$ increases from 0$^{\circ}$ (parallel) to 90$^{\circ}$ (perpendicular), these absorption peaks decrease in intensity.

The spectrum for perpendicular polarization ($\beta$ = 90$^{\circ}$) shows a new peak around 1.9\,eV, which we assign to the $E_{12}/E_{21}$ transition. As stated above, this transition is expected for light polarized perpendicular to the nanotube axis (Fig.\,\ref{I_schematic}). A closer look at the polarization-dependent spectra allowed us to identify this peak in all spectra for polarization angles equal to or larger than 60$^{\circ}$.  Furthermore, it should be noted that this peak exists even in the suspension spectrum shown in Fig.\,\ref{II_suspension_spectrum}, although peak assignment was impossible since the nanotubes in the suspension are randomly oriented.

Figures \ref{III_spectra}(b)--(d) compare the 0$^{\circ}$ ($A_\parallel$) and 90$^{\circ}$ ($A_\perp$) spectra in more detail. In these figures, a polynomial baseline was subtracted from each spectrum; see Sec.\,IV for more details about this procedure. In Fig.\,\ref{III_spectra}(b), the red and black curves represent $A_\parallel$ and $A_\perp$, respectively, where the $A_\perp$ spectrum is multiplied by 3.2 so that the $E_{11}$ peak coincides in intensity between the two spectra.  As a result, the two spectra deviate from each other only in the spectral region of the $E_{12}/E_{21}$ peak.  In Fig.\,\ref{III_spectra}(c), $A_\perp$ multiplied by 3.2  is plotted in the upper ($y > 0$) plane, whereas $A_\parallel$ is plotted in the lower ($y < 0$) plane. The vertical dashed lines indicate the positions of the $E_{11}$ peak, the $E_{11}$ phonon sideband peak, the $E_{12}/E_{21}$ peak, and the $E_{22}$ peak, respectively. The blue curve is $3.2A_\perp-A_\parallel$, which is essentially zero everywhere except for the $E_{12}/E_{21}$ feature since the $E_{12}/E_{21}$ feature only appears in $A_\perp$. Finally, Fig.\,\ref{III_spectra}(d) shows a spectral difference ($3.2A_\perp-A_\parallel$) normalized by $A_\parallel$. In this spectrum, the effects of the $E_{11}$ peak, the $E_{11}$ phonon sideband peak, and the $E_{22}$ peak are nearly eliminated, leaving a pronounced single peak due to the $E_{12}/E_{21}$ exciton.

\section{SPECTRAL ANALYSIS}
To extract quantitative information from the obtained polarization-dependent spectra, we performed spectral analysis. We fit each spectrum with a function consisting of Lorentzians representing the absorption peaks and a polynomial function representing the baseline:
\begin{equation}\label{eq_fitting}
A\equiv -\ln(T)=\sum^{3\,\text{or}\,4}_{n=1}a_n\frac{(b_n/2)^2}{(E_\text{ph}-c_n)^2+(b_n/2)^2}+\sum_{m=0}^{4}d_mE_\text{ph}^m,
\end{equation}
where $E_\text{ph}$ is the photon energy, acting as the independent variable, and $a_n$, $b_n$, $c_n$, and, $d_m$ are the fitting parameters. $a_n$, $b_n$, and $c_n$ are the peak amplitude, full width at half maximum, and peak position, respectively, of the $n$-th peak, while $d_m$ is the $m$-th polynomial coefficient. We considered polynomials of order up to $m = 4$. We performed fitting on all spectra with polarization angles from $ -5^{\circ}$ to 90$^{\circ}$ with a step size of $5^{\circ}$. The spectra from $ -5^{\circ}$ to 30$^{\circ}$ were fit with a polynomial function and three Lorentzians, to take account of the $E_{11}$ peak, the $E_{11}$ phonon sideband peak, and the $E_{22}$ peak. The spectra from 35$^{\circ}$ to 90$^{\circ}$ were fit with four Lorentzians to take into account the $E_{12}/E_{21}$ peak as well.

Figure \ref{IV_fit} shows fitting results for the spectra for $\beta =$ 0$^{\circ}$, 45$^{\circ}$, 60$^{\circ}$, and 90$^{\circ}$. The solid black lines are experimental data. The dashed red lines indicate the overall fit functions. The blue curves indicate the individual components of the fit function. 
Note that the spectrum for 0$^{\circ}$ shown in Fig.\,\ref{IV_fit}(a) does not contain the $E_{12}/E_{21}$ peak.

Figures \ref{IV_fit_individual}(a)-\ref{IV_fit_individual}(d) plot the extracted polarization-dependent spectra for the $E_{11}$ peak, the $E_{11}$ phonon sideband peak, the $E_{12}/E_{21}$ peak, the $E_{22}$ peak, and the polynomial baseline, respectively. The shape of the baseline slightly changes with the polarization angle. As the angle increases, the overall intensities of the baseline, the $E_{11}$ peak, the $E_{11}$ phonon sideband, and the $E_{22}$ peak decrease, while the $E_{12}/E_{21}$ peak grows in intensity. The peak widths of the $E_{11}$ and $E_{22}$ peaks are $\sim$120\,meV and $\sim$190\,meV, respectively. 

Finally, Figs.\,\ref{IV_peak_integral}(a)-\ref{IV_peak_integral}(d) plot the integrated peak intensities of the $E_{11}$ peak, the $E_{11}$ phonon sideband, the $E_{22}$ peak, the $E_{12}/E_{21}$ peak, respectively, as a function of polarization angle $\beta$. While the integrated intensities of the $E_{11}$ peak, the $E_{11}$ phonon sideband, and the $E_{22}$ peak decrease as the polarization angle $\beta$ increases, the integrated intensity of the $E_{12}/E_{21}$ peak increases.

\section{DISCUSSION}

\subsection{Nematic order parameter}

Since the average length of SWCNTs ($\sim$200\,nm) is much larger than the film thickness ($<$10\,nm) in our sample, we use the two-dimensional (2D) theory of the optical absorption by an ensemble of anisotropic molecules, described in Appendix \ref{Appendix:Theory} Section 2, to discuss our experimental data.  We assume that the nanotubes' angular distribution $f(\theta)$ can be represented by the following Gaussian function with $\theta=0$ as the alignment direction:
\begin{equation}\label{eq_f_gauss}
f(\theta)=\frac{1}{\text{erf}\left( \pi/\sqrt{2}\sigma\right) \sqrt{2\pi\sigma^2}}\left( e^{-\frac{\theta^2}{2\sigma^{2}}}+e^{-\frac{(\theta-\pi)^2}{2\sigma^{2}}}\right),
\end{equation}
where $\theta$ is the angle between the macroscopic alignment direction and an individual nanotube and $\sigma$ is the standard deviation.  Note that the nanotubes are distributed in an angular range of $0 \leq \theta \leq \pi$, and $f(\theta)$ is normalized in this range, i.e., $\int_{0}^{\pi}f(\theta)d\theta=1$. 

\begin{figure}
	\includegraphics[width=\linewidth]{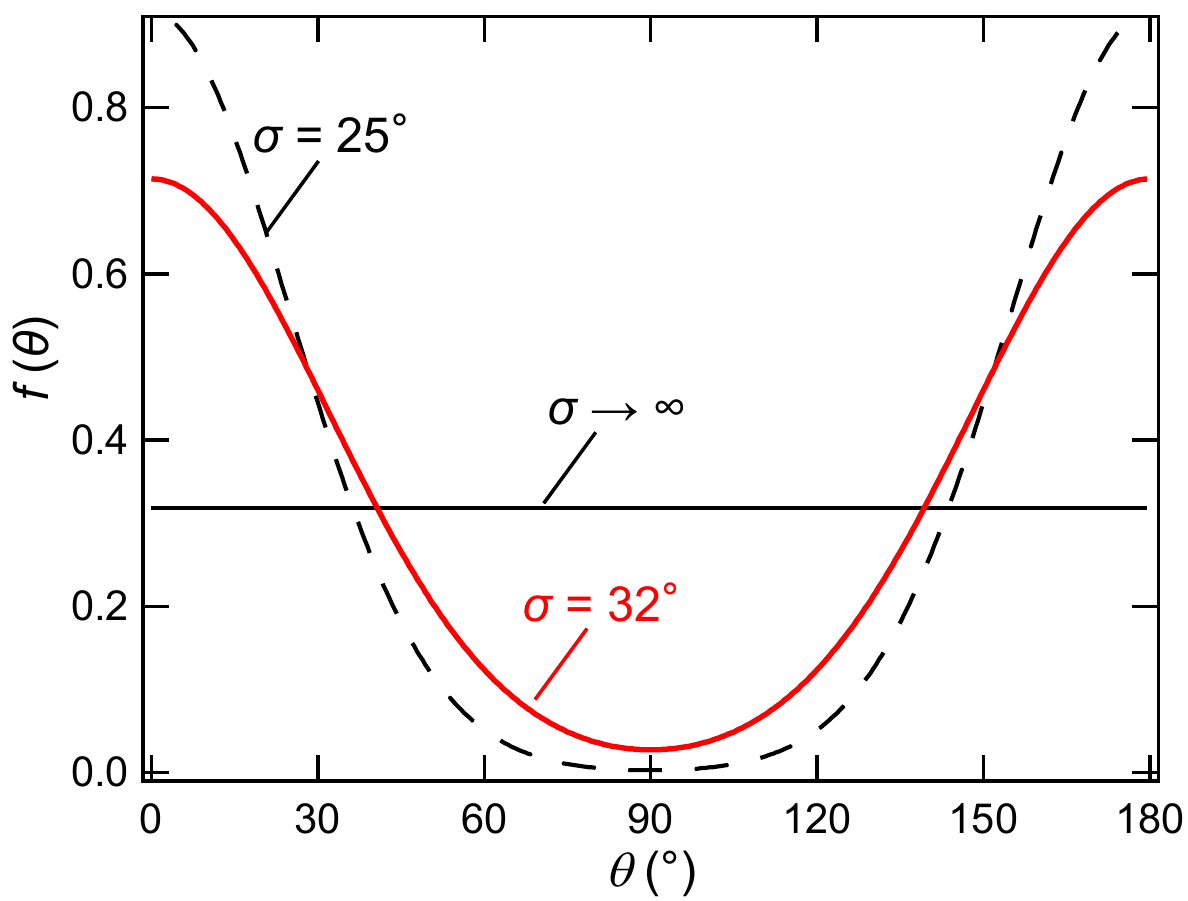}
	\caption{Simulated nanotubes' angular distribution $f(\theta)$, based on Eq.\,(\ref{eq_f_gauss}). The three traces correspond to $\sigma = 25^\circ$, $\sigma = 32^\circ$, and $\sigma \rightarrow \infty$, respectively.}
	\label{V_ftheta}
\end{figure}

Figure \ref{V_ftheta} shows three examples of $f(\theta)$ for the cases of $\sigma$ = 25$^\circ$, 32$^\circ$, and $\infty$. When $\sigma=25^\circ$ (shown as a black dashed line), clear alignment along $\theta=0$ is observed. As $\sigma$ increases, the distribution function $f\left( \theta\right) $ becomes flatter. Finally, when $\sigma \rightarrow \infty$, $f\left( \theta\right) \rightarrow 1/\pi$ as indicated by the black solid line. 

\begin{figure}
	\includegraphics[width=\linewidth]{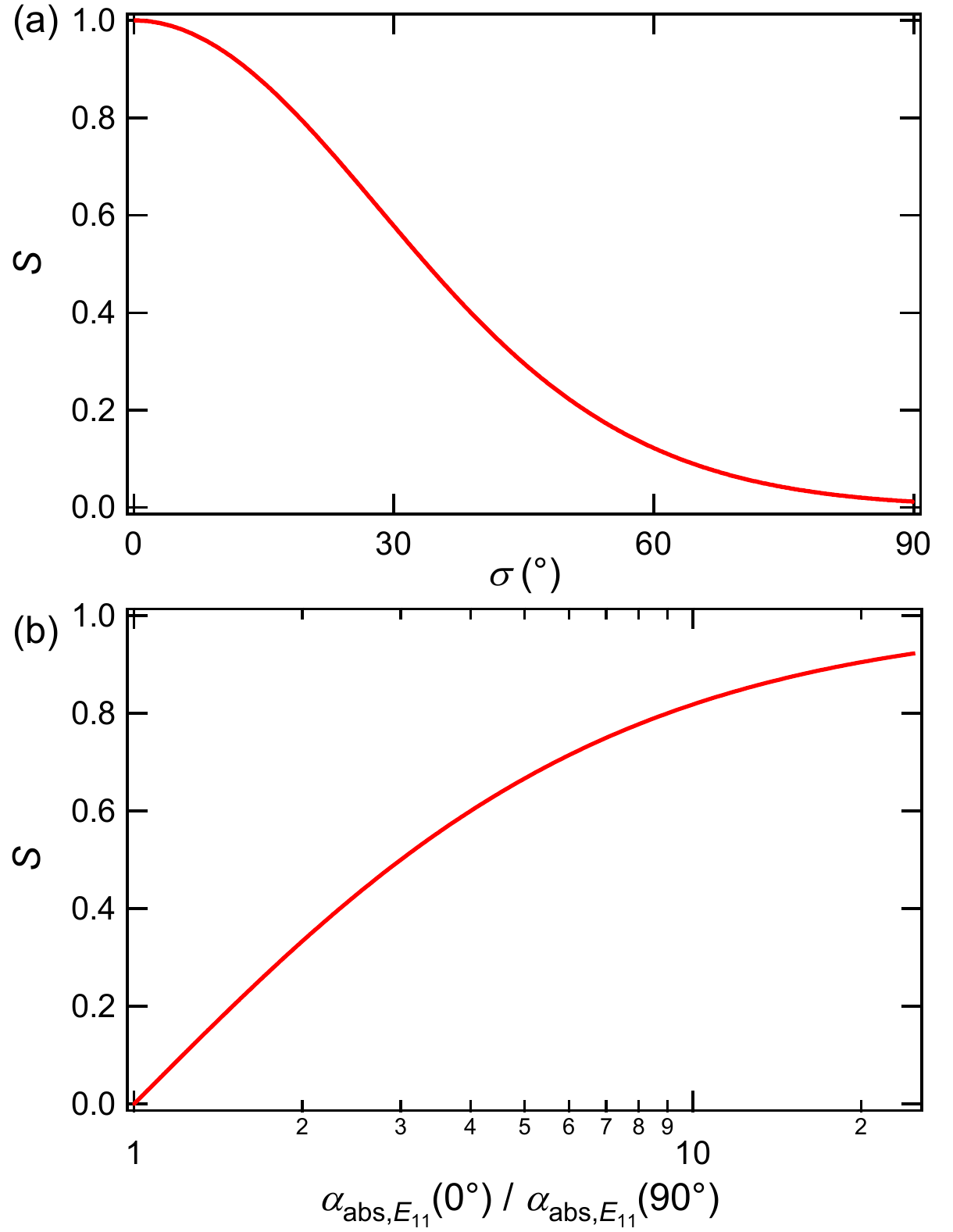}
	\caption{(a)~Nematic order parameter $S$ as a function of standard deviation angle $\sigma$ based on Eq.\,(\ref{eq_order_para_f}). (b)~Nematic order parameter $S$ as a function of absorption ratio between parallel and perpendicular polarization based on Eq.\,(\ref{eq_attenuation_ratio}).}
	\label{V_simulated_S_ratio}
\end{figure}

With the distribution function $f\left( \theta\right)$ given by Eq.\,(\ref{eq_f_gauss}), the 2D order parameter $S$, defined by Eq.\,(\ref{eq_order_para_2D}), can be calculated as
\begin{align}\label{eq_order_para_f}
S & =\int_{0}^{\pi}f(\theta)\left( 2\cos^2\theta-1\right)  d\theta \nonumber \\
& =\frac{e^{-2\sigma^2}}{2\text{erf}(\pi/\sqrt{2}\sigma)} \left[ \text{erf}\left( \frac{\pi}{\sqrt{2}\sigma}-i\sqrt{2}\sigma\right) \right.\nonumber \\
& \left. + \text{erf}\left( \frac{\pi}{\sqrt{2}\sigma}+i\sqrt{2}\sigma\right) \right] .
\end{align}
Since $S$ and $\sigma$ have one-to-one correspondence, we can plot $S$ as a function of $\sigma$, as shown in Fig.\,\ref{V_simulated_S_ratio}(a). When $\sigma \rightarrow 0$, $S \rightarrow 1$, as expected. As $\sigma$ increases, $S$ monotonically decreases, and finally, when $\sigma \rightarrow \infty$, $S \rightarrow 0$.

When the polarization angle is $\beta$ with respect to the nanotube alignment direction (see Fig.\,\ref{II_setup}), the absorption coefficient for incident light with photon energy $E_\text{ph}$ is given by
\begin{align}\label{eq_attenuation}
\alpha_\text{abs}(\beta)& = \frac{NE_\text{ph}}{\hbar c n_0}\left( \alpha_1''\frac{\int_{0}^{\pi}f(\theta)\cos^2(\theta-\beta)d\theta}{\int_{0}^{\pi}f(\theta)d\theta} \right. \nonumber \\
&+ \left. \alpha_2''\frac{\int_{0}^{\pi}f(\theta)\sin^2(\theta-\beta)d\theta}{\int_{0}^{\pi}f(\theta)d\theta}\right) \nonumber \\
& = \frac{NE_\text{ph}}{\hbar c n_0}\left(  \alpha_1''\int_{0}^{\pi}f(\theta-\beta)\cos^2(\theta-\beta) d\theta \right. \nonumber \\
&\left.  +\alpha_2''\int_{0}^{\pi}f(\theta-\beta)\sin^2\left (\theta-\beta\right ) d\theta\right),
\end{align}
where $N$ is the total number of SWCNTs, $\hbar$ is the reduced Planck constant, $c$ is the speed of light, $n_0$ is the refractive index, and $\alpha_1''$ ($\alpha_2''$) is the imaginary part of the molecular polarizability, $\alpha$, of an individual SWCNT parallel (perpendicular) to the tube axis.  See Appendix~\ref{Appendix:Theory} for more details.

We assume that the polarizability of an $E_{ii}$ transition is parallel to the nanotube axis ($\xi_\text{2D} = 0^\circ$) whereas that of an $E_{ij}$ ($i\ne j$) transition is perpendicular to the nanotube axis ($\xi_\text{2D} = 90^\circ$), where $\xi_\text{2D} = \tan^{-1}\left ( \sqrt{\alpha_1^{''}/\alpha_2^{''}} \right )$ (see Appendix~\ref{Appendix:Theory} Section 2). Namely, to consider the $E_{11}$ transition, we assume $\xi_\text{2D}=0^\circ$, i.e., $\alpha_1'' \ne 0$ and $\alpha_2'' = 0$. With the distribution $f(\theta)$ given by Eq.\,(\ref{eq_f_gauss}), the absorption coefficient for the $E_{11}$ transition becomes
\begin{align}\label{eq_attenuation_E11}
\alpha_{\text{abs},E_{11}}(\beta)& = \frac{NE_{11}}{\hbar c n_0}\alpha_1''\int_{0}^{\pi}f(\theta-\beta)\cos^2(\theta-\beta) d\theta \nonumber \\ 
& = \frac{NE_{11}}{\hbar c n_0}\alpha_1''\left[ \frac{1}{2} +\right.\nonumber \\ 
& \frac{e^{-2\sigma^2}}{4\text{erf}(\pi/\sqrt{2}\sigma)}\left \{ \text{erf}\left( \frac{\pi}{\sqrt{2}\sigma}-i\sqrt{2}\sigma \right) \right. \nonumber \\ 
& \left. \left. +\text{erf}\left( \frac{\pi}{\sqrt{2}\sigma}+i\sqrt{2}\sigma\right) \right \} \cos2\beta \right].
\end{align}
Similarly, by assuming that $\xi_\text{2D}=90^\circ$, we obtain the absorption coefficient for the $E_{12}/E_{21}$ transition as
\begin{align}\label{eq_attenuation_E12}
\alpha_{\text{abs},E_{12}}(\beta)& = \frac{NE_{12}}{\hbar c n_0}\alpha_2''\int_{0}^{\pi}f(\theta-\beta)\sin^2(\theta-\beta) d\theta. \nonumber \\ 
& = \frac{NE_{12}}{\hbar c n_0}\alpha_2''\left[ \frac{1}{2} + \right.\nonumber \\ 
& \frac{e^{-2\sigma^2}}{4\text{erf}(\pi/\sqrt{2}\sigma)}\left \{ \text{erf}\left( \frac{\pi}{\sqrt{2}\sigma}-i\sqrt{2}\sigma \right) \right. \nonumber \\ 
& \left. \left. -\text{erf}\left( \frac{\pi}{\sqrt{2}\sigma}+i\sqrt{2}\sigma\right) \right \} \cos2\beta \right].
\end{align}

Therefore, when the light polarization is parallel to the macroscopic alignment direction of the film, the absorption coefficient of the $E_{11}$ transition is given by
\begin{align}\label{eq_attenuation_par}
\alpha_{\text{abs},E_{11}}(0^\circ) & = \frac{NE_{11}}{\hbar c n_0}\alpha_1''\int_{0}^{\pi}f(\theta)\cos^2(\theta) d\theta \nonumber \\
& = \frac{NE_{11}}{\hbar c n_0}\frac{1+S}{2}\alpha_1''.
\end{align}
On the other hand, when the light polarization is perpendicular to the alignment direction, the absorption coefficient of the $E_{11}$ transition is given by
\begin{align}\label{eq_attenuation_per}
\alpha_{\text{abs},E_{11}}(90^\circ) & = \frac{NE_{11}}{\hbar c n_0}\alpha_1''\int_{0}^{\pi}f\left (\theta-\frac{\pi}{2}\right )\cos^2\left (\theta-\frac{\pi}{2} \right ) d\theta\nonumber \\
& = \frac{NE_{11}}{\hbar c n_0}\frac{1-S}{2}\alpha_1''.
\end{align}
Hence, the absorption coefficient ratio between parallel and perpendicular polarization is given by
\begin{equation}\label{eq_attenuation_ratio}
\frac{\alpha_{\text{abs},E_{11}}(0^\circ)}{\alpha_{\text{abs},E_{11}}(90^\circ)} =\frac{1+S}{1-S}.
\end{equation}
By reversing Eq.\,(\ref{eq_attenuation_ratio}), we can express $S$ in terms of the absorption coefficient ratio as
\begin{equation}\label{eq_S_attenuation_ratio}
S =\frac{\alpha_{\text{abs},E_{11}}(0^\circ)/\alpha_{\text{abs},E_{11}}(90^\circ) -1}{\alpha_{\text{abs},E_{11}}(0^\circ)/\alpha_{\text{abs},E_{11}}(90^\circ)+1}.
\end{equation}
In Fig.\,\ref{V_simulated_S_ratio}(b), $S$ is plotted as a function of $\alpha_{\text{abs},E_{11}}(0^\circ)/\alpha_{\text{abs},E_{11}}(90^\circ)$. When $\alpha_{\text{abs},E_{11}}(0^\circ)/\alpha_{\text{abs},E_{11}}(90^\circ)=1$, there is no anisotropy, meaning that $S = 0$.  As the absorption ratio increases, $S$ increases. As $\alpha_{\text{abs},E_{11}}(0^\circ)/\alpha_{\text{abs},E_{11}}(90^\circ) \rightarrow \infty$, $S$ asymptotically approaches 1. 
\begin{figure}
	\includegraphics[width=\linewidth]{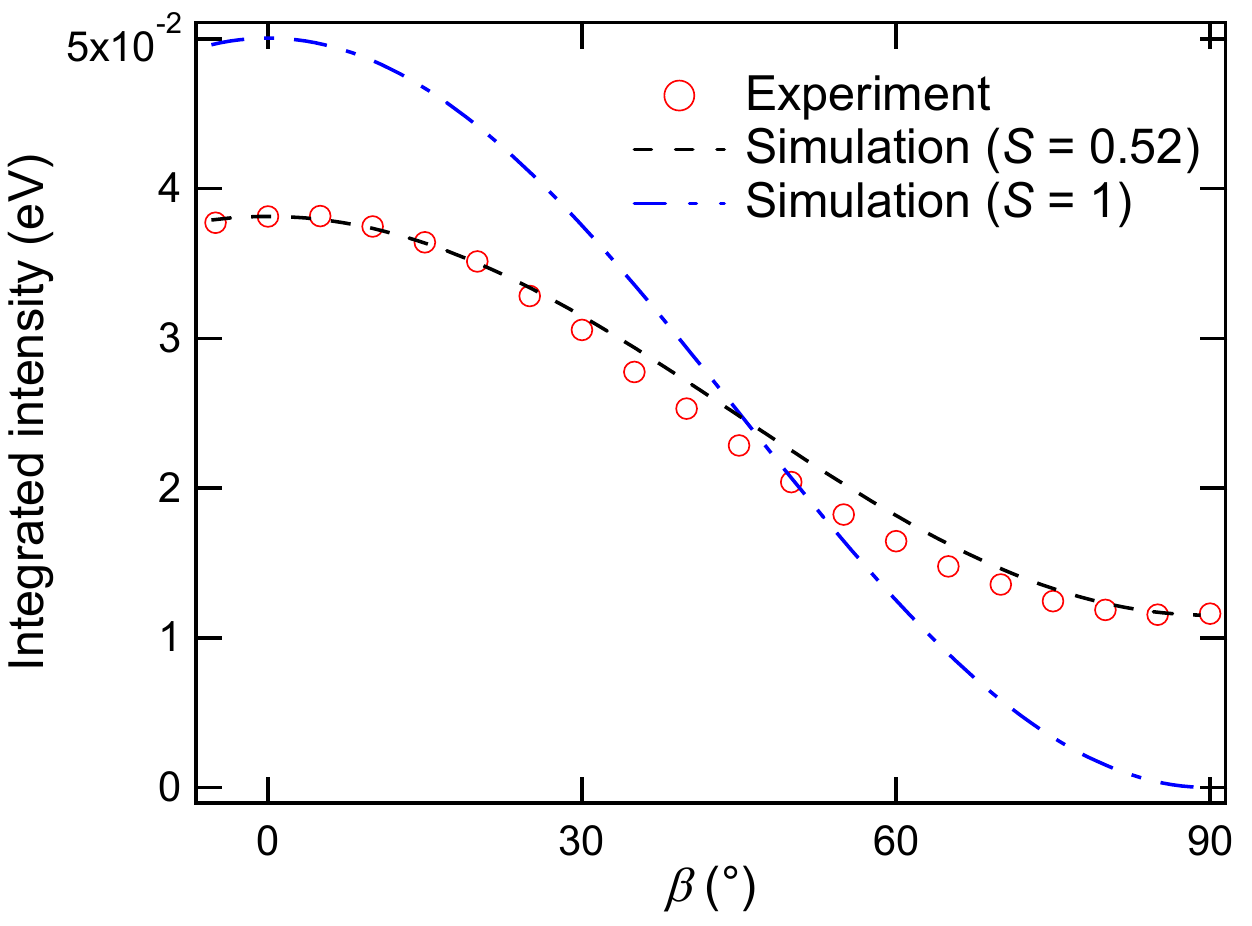}
	\caption{Polarization angle dependence of the integrated intensity of the $E_{11}$ peak.  Black dashed curve: theoretical calculation assuming $S = 0.52$.  Red open circles: experimental data.  The experimental observation is well reproduced by the theoretical curve.  Blue dash-dotted curve: theoretical calculation assuming perfect alignment, i.e., $S = 1$.}
	\label{V_simulated_integral}
\end{figure}

\subsection{Angular dependence of $\boldsymbol{E_{11}}$ and $\boldsymbol{E_{12}/E_{21}}$ absorption intensities}
When the reflection loss can be neglected, the quantity we measured experimentally, i.e., the attenuation $A = -\ln{(T)}$ is directly proportional to the absorption coefficient.  Namely, $A = \alpha_\text{abs}l$, where $l$ is the film thickness.
%
Therefore, the experimentally determined $E_{11}$ integrated peak intensity ratio ($A_\parallel / A_\perp$) can be assumed to be equal to $\alpha_{e,E_{11}}(0^\circ)/\alpha_{e,E_{11}}(90^\circ)$.  From Fig.\,\ref{IV_peak_integral}(a), $A_\parallel / A_\perp$ is determined to be 3.2, which, according to the plot in Fig.\,\ref{V_simulated_S_ratio}(b), corresponds to $S =$ 0.52.  Accordingly, from Fig.\,\ref{V_simulated_S_ratio}(a) and Eq.\,(\ref{eq_order_para_f}), $\sigma$ is determined to be 32$^\circ$. Figure \ref{V_ftheta} plots the angular distribution of nanotubes for this case as a red solid curve. 

Furthermore, we calculated the integrated intensity of the $E_{11}$ peak in absorption coefficient as a function of polarization angle $\beta$ for $S = 0.52$, as shown in Fig.\,\ref{V_simulated_integral} as a black dashed line together with the experimental data (red open circles).  The calculated values are normalized by the experimental value for 0$^\circ$. The observed angular dependence is accurately reproduced by the theoretical curve, confirming the overall correctness of our theoretical analysis.  Finally, the blue dash dotted line in Fig.\,\ref{V_simulated_integral} represents the angular dependence of the $E_{11}$ integrated absorption intensity calculated assuming perfect alignment, i.e., $S = 1$.

\subsection{Energy and oscillator strength of the $\boldsymbol{E_{12}/E_{21}}$ transition}
\begin{figure}
	\includegraphics[width=\linewidth]{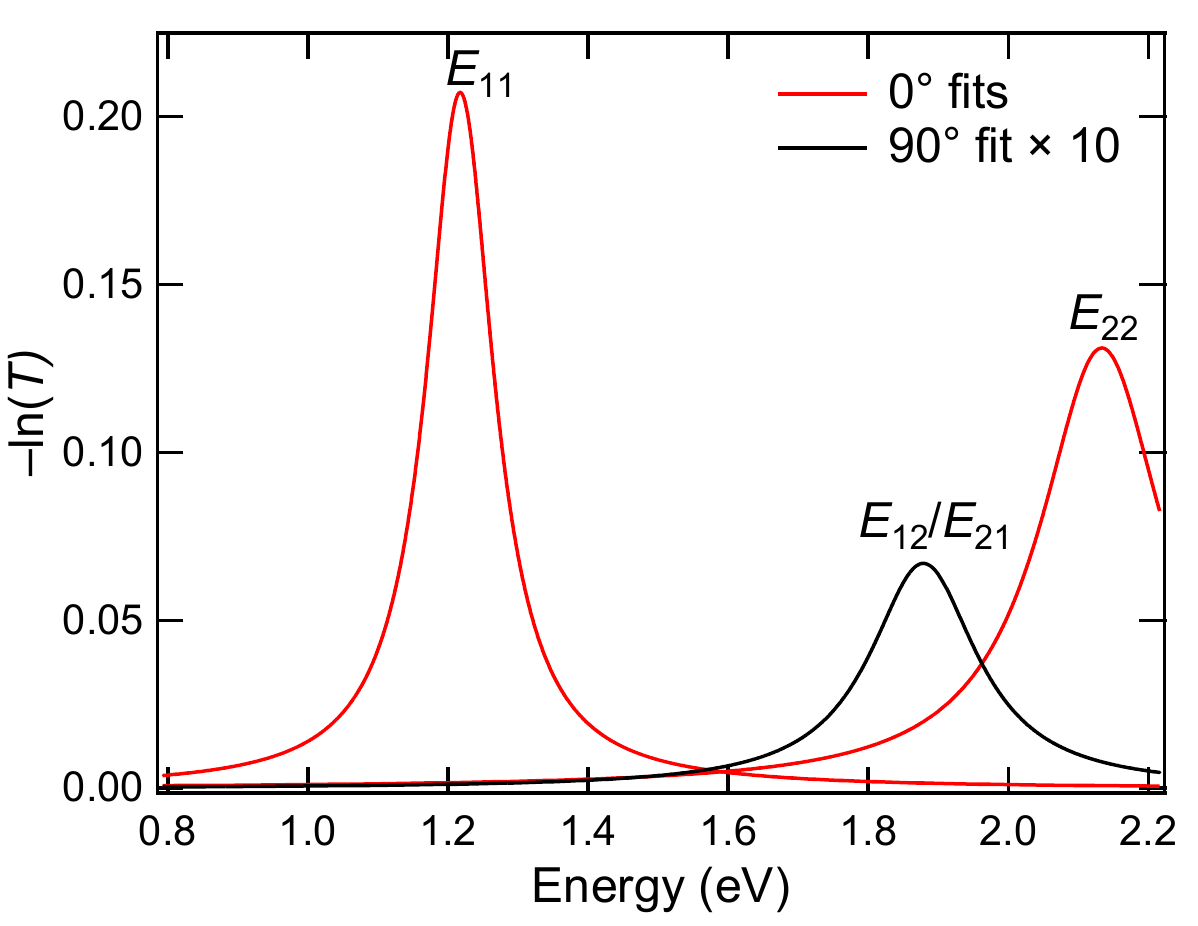}
	\caption{Fit peak comparison of $E_{11}$ and $E_{22}$ for $0^\circ$, and $E_{12}/E_{21}$ for $90^\circ$. The $E_{12}/E_{21}$ peak is multiplied by 10.}
	\label{V_fit}
\end{figure}

Figure \ref{V_fit} shows a parallel-polarization spectrum ($\beta = 0^\circ$) exhibiting the $E_{11}$ and $E_{22}$ peaks, together with a perpendicular-polarization spectrum ($\beta = 90^\circ$) exhibiting the $E_{12}/E_{21}$ peak, which were extracted from the raw experimental data through the spectral analysis described in Section IV.  The perpendicular-polarization spectrum was multiplied by 10.  The energy position of the $E_{12}/E_{21}$ peak is 1.88\,eV, which is 1.54 times that of the $E_{11}$ peak (1.22\,eV) and 0.88 times that of the $E_{22}$ peak (2.13\,eV). Previously, the $E_{12}/E_{21}$ transition was observed through cross-polarized photoluminescence excitation experiments~\cite{ChuangetAl08PRB} and circular dichroism measurements~\cite{WeietAl16NC,AoetAl16JACS}. The reported energies range from 1.88 to 1.93\,eV. These fluctuations can be attributed to the different dielectric constants of the surrounding of the nanotubes studied under different conditions~\cite{UryuAndo06PRB,UryuAndo12PRB}. 
Uryu and Ando calculated the energies of the $E_{11}$, $E_{12}/E_{21}$, and $E_{22}$ peaks for SWCNTs as a function of dielectric constant $\kappa$ and diameter~\cite{UryuAndo06PRB}. While we found no single value of $\kappa$ that simultaneously makes the three calculated energies match the experimental values, we found reasonable overall agreement when $1.8 < \kappa < 3.5$.

We next discuss the oscillator strength ratio of the $E_{12}/E_{21}$  and $E_{11}$ transitions. Directly from the traces presented in Fig.\,\ref{V_fit}, we can determine this ratio to be $I_{12}/I_{11}$ = 0.05.  Here, $I_{11}$ ($I_{12}$) is the integrated intensity of the $E_{11}$ ($E_{12}/E_{21}$) peak in the parallel-polarization (perpendicular-polarization) spectrum.  It is important to note that this ratio is independent of $S$.  This can be easily seen by comparing Eq.\,(\ref{eq_attenuation_par}) and
\begin{align}\label{eq_d_vc_2}
\alpha_{\text{abs},E_{12}}(90^\circ) & = \frac{NE_{12}}{\hbar c n_0}\alpha_2''\int_{0}^{\pi}f \left (\theta-\frac{\pi}{2}\right )\sin^2\left (\theta-\frac{\pi}{2} \right) d\theta\nonumber \\
& = \frac{NE_{12}}{\hbar c n_0}\frac{1+S}{2}\alpha_2''.
\end{align}
Namely,
\begin{align}\label{eq_osc_strength_2}
\frac{\alpha_{\text{abs},E_{11}}(0^\circ)}{\alpha_{\text{abs},E_{12}}(90^\circ)} = \frac{E_{11}\alpha_1''}{E_{12}\alpha_2''}.
\end{align}
By equating this ratio to $I_{11}/I_{12}$, we can also obtain the ratio of the imaginary part of the molecular polarizability for perpendicular polarization at $E_{12}$ to that for parallel polarization at $E_{11}$
\begin{align}\label{eq_osc_strength_2}
\frac{\alpha_2''}{\alpha_1''} = \frac{E_{11}}{E_{12}} \times 0.05 = 0.03.
\end{align}

Finally, we can also use the obtained value of $I_{12}/I_{11}$ = 0.05 to get a value for the dielectric constant, $\kappa$, through comparison with the theoretical calculations of this ratio by Uryu and Ando~\cite{UryuAndo06PRB}.  The radiation power absorbed by a nanotube can be expressed as
\begin{align}\label{UryuPower}
P_\parallel = \frac{1}{2} \sigma_{11}'D^2 \\
P_\perp = \frac{1}{4} \sigma_{12}'D^2
\end{align}
for parallel and perpendicular polarizations, respectively.  Here, $\sigma_{11}'$ ($\sigma_{12}'$) is the real part of the optical conductivity parallel (perpendicular) to the nanotube axis at $E_\text{ph} = E_{11}$ ($E_\text{ph} = E_{12}$) and $D$ is the amplitude of the electric field of light. Note that these expressions take into account the fact that only the wavenumber components $\pm2\pi/L$ (where $L$ is the nanotube circumference) of the incident light can excite the $E_{12}/E_{21}$ transition whereas only the zero-wavenumber component of the incident light can excite the $E_{11}$ transition; the inclusion of the $\pm2\pi/L$ components corresponds to the simultaneous excitation of the $E_{12}$ and $E_{21}$ transitions~\cite{AjikiAndo94Physica}.   Spectrally integrated and properly normalized values of $\sigma_{12}'$ and $\sigma_{11}'$ (and thus those of $2P_\perp$ and $P_\parallel$) can be found in Fig.\,7 of Ref.\,\cite{UryuAndo06PRB}. Hence, we compared the calculated ratio $2P_\perp/P_\parallel$ with our experimental value $2I_{12}/I_{11}$ = 0.10 and obtained $\kappa$ = 1.52. This value is slightly outside the range we deduced from the peak energy consideration above ($1.8 < \kappa < 3.5$).  A better treatment of the surrounding dielectrics~\cite{UryuAndo12PRB} as well as inclusion of higher-order terms in the band structure calculation are needed to fully explain the experimental results quantitatively.  

\section{Summary}
We prepared a macroscopic film of highly aligned single-chirality (6,5) SWCNTs and performed a polarization-dependent optical absorption spectroscopy study.  In addition to the usual $E_{11}$ and $E_{22}$ exciton peaks for parallel-polarized light, we observed a clear absorption peak due to the $E_{12}$/$E_{21}$ exciton peak for perpendicular-polarized light.  Unlike previous observations of cross-polarized excitons in polarization-dependent photoluminescence and circular dichroism spectroscopy experiments, our direct absorption observation allowed us to quantitatively analyze this resonance.  We determined the energy of this peak to be 1.54 times that of the $E_{11}$ peak and the oscillator strength of this resonance to be 0.05 times that of the $E_{11}$ peak.  These values, in light of theoretical calculations available in the literature, led to an assessment of the environmental effect on the strength of Coulomb interactions in this aligned single-chirality SWCNT film.

\section*{Acknowledgements}
We thank Seiji Uryu, Tsuneya Ando, and Katsumasa Yoshioka for useful discussions.  This work was supported by the U.S.\ Department of Energy Basic Energy Sciences through grant no.\ DEFG02-06ER46308 (optical spectroscopy experiments), the U.S.\ National Science Foundation through award no.\ ECCS-1708315 (modeling), and the Robert A.\ Welch Foundation through grant no.\ C-1509 (sample preparation). K.Y.\ acknowledges support by JSPS KAKENHI through Grant Numbers JP16H00919, JP17K14088, JP25107003, JP17H01069, JP17H06124, and JP15K21722, JST CREST through Grant Number JPMJCR17I5, Japan, and the Yamada Science Foundation.

\newpage

\appendix

\section{Chirality Purity Determination}
\label{Appendix:Chirality}

\begin{figure}[htb]
	\includegraphics[width=\linewidth]{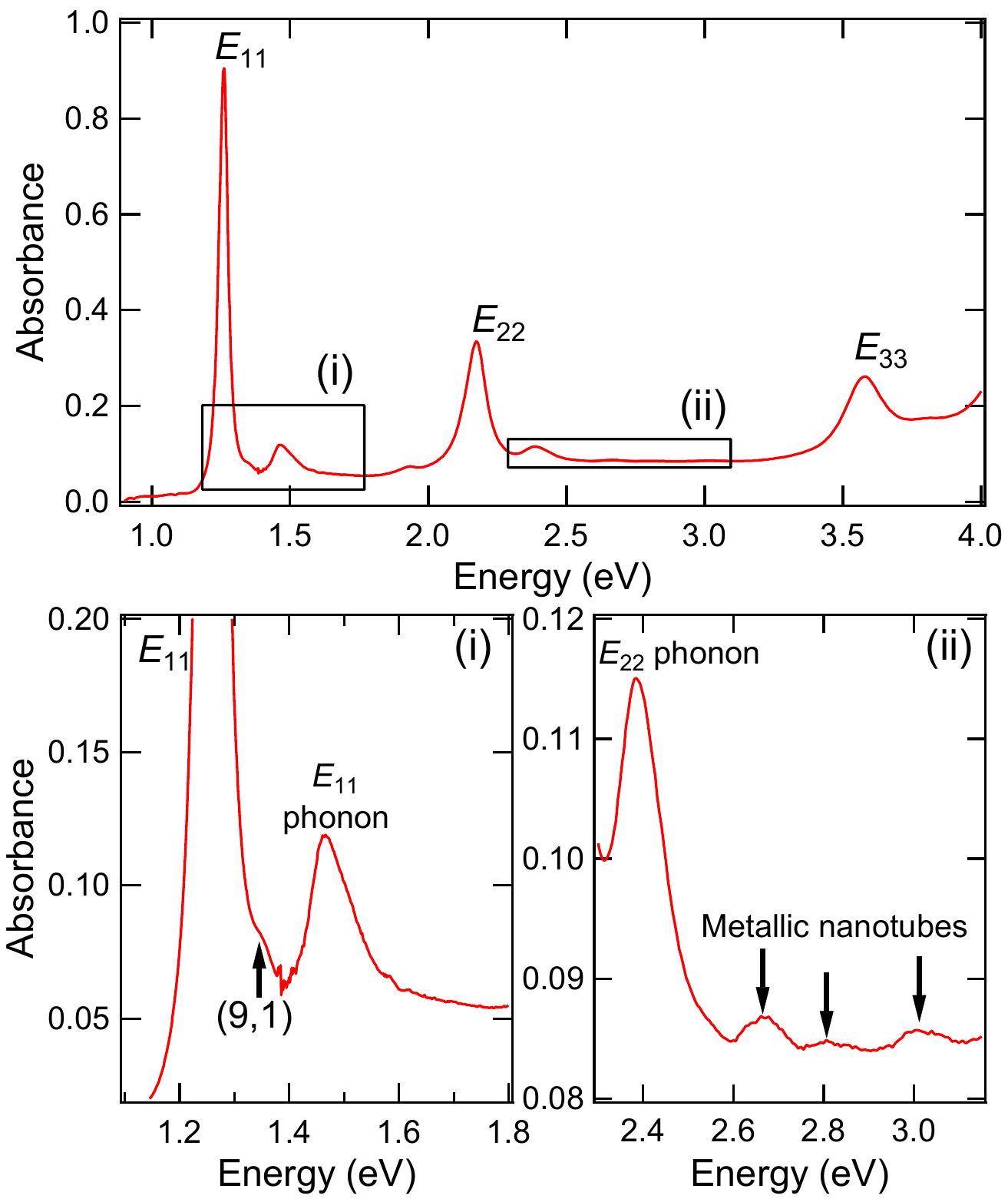}
	\caption{Absorbance spectrum for the SWCNT suspension used for making the film studied in this study.  Two spectral regions of interest -- (i) and (ii) -- are expanded in the bottom two panels.}
	\label{Residual}
\end{figure}

\begin{figure}[htb]
	\includegraphics[width=\linewidth]{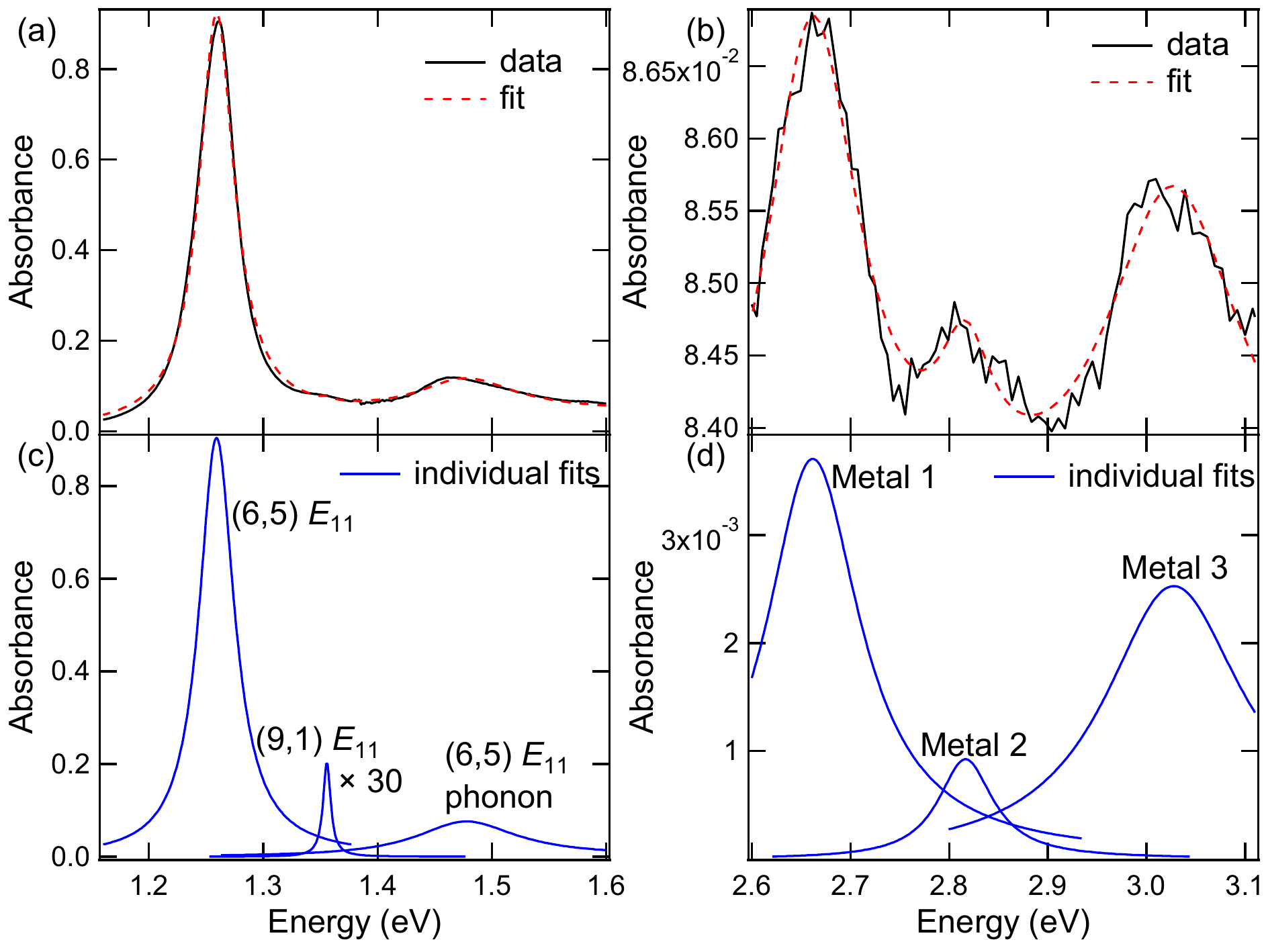}
	\caption{Spectral fitting analysis performed to determine the relative peak intensities of the $E_{11}$ peaks of (6,5), (9,1), and metallic SWCNTs in the sample.}
	\label{KatauraFits}
\end{figure}

To assess the chirality purity of our sample quantitatively, we analyzed the absorption spectrum shown in Fig.\,\ref{II_suspension_spectrum} using the method described in Ref.\,\cite{LiuetAl11NC}.  The spectrum is reproduced in Fig.\,\ref{Residual} with two spectral regions of interest expanded.  In Region (i), we observe a shoulder, which we attribute to the $E_{11}$ peak of residual (9,1) SWCNTs.  In Region (ii), there are three small peaks, which can be attributed to the $E_{11}$ peaks of metallic SWCNTs.  Through line-fitting analysis shown in Fig.\,\ref{KatauraFits}, we determined the relative peak intensities of the observed peaks, as summarized in Table~\ref{Table}.  From these values, neglecting any ($n$,$m$) dependence of oscillator strength, we can calculate the relative population of (6,5) SWCNTs to be (41.658/41.941)$\times$100 = 99.3\%.
\begin{table}[ht]
\caption{Relative integrated peak intensities of the $E_{11}$ peaks of (6,5), (9,1), and metallic SWCNTs in the sample.}
\begin{center}
\begin{tabular}{|c||c|c|c|c|c||c|}
\hline
Chirality & (6,5) & (9,1) & Metal 1 & Metal 2 & Metal 3 & Total\\
\hline\hline
Area & 41.658 & 0.058 & 0.093 & 0.015 & 0.117 & 41.941\\
\hline
\% & 99.33 & 0.13 & 0.22 & 0.04 & 0.27 & 100\\
\hline
\end{tabular}
\end{center}
\label{Table}
\end{table}%

\section{Optical Absorption and Nematic Order Parameter of an Ensemble of Anisotropic Molecules}
\label{Appendix:Theory}
\subsection{Three-dimensional (3D) case}
\begin{figure}
	\includegraphics[width=\linewidth]{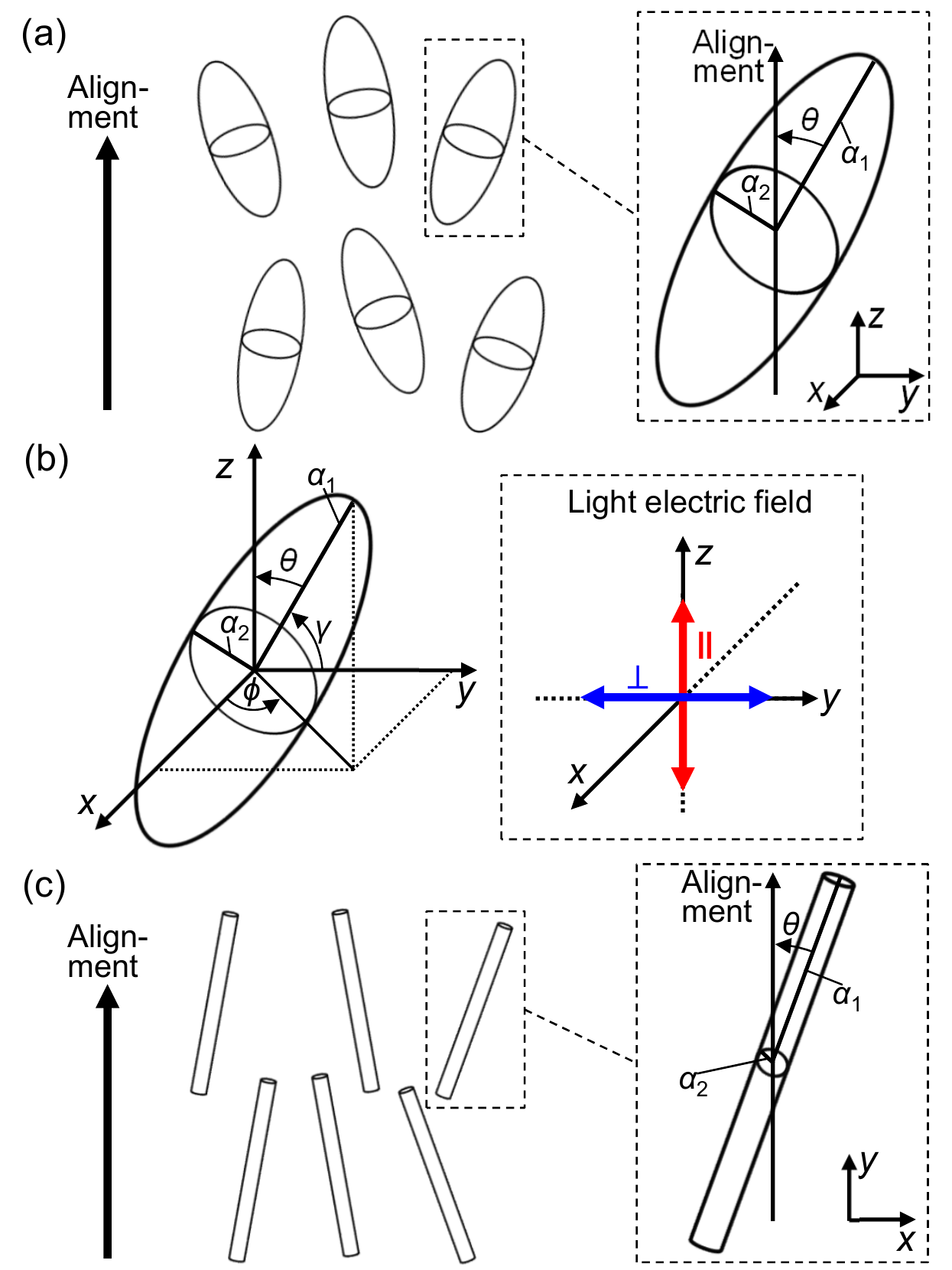}
	\caption{(a)~An ensemble of spheroidal molecules in 3D space. (b)~Detailed illustration of a molecule in a 3D global coordinate system. The alignment direction is along the $z$-axis. (c)~Schematic of a 2D ensemble of carbon nanotubes. The alignment direction is along the $y$-axis.}
	\label{V_molecule}
\end{figure}

Let us cosnider an ensemble of spheroidal molecules and their anisotropic optical absorption properties.  As shown in Fig.\,\ref{V_molecule}(a), we define the molecular polarizability along the long axis as $\alpha_1$ and the molecular polarizability along the short axis as $\alpha_2$. $\theta$ is the angle between the alignment direction of the ensemble and the long axis of the particular individual molecule that we examine. 

When an electric field is applied parallel to the alignment direction (which is the $z$-direction in Fig.\,\ref{V_molecule}(a)), the expectation value (i.e., the ensemble average) of the molecular polarizability $\langle\alpha\rangle_{\text{3D}}$ is given by
\begin{align}\label{eq_alpha_par}
\langle\alpha\rangle_{\parallel,\text{3D}} & =\alpha_1\langle\cos^2\theta\rangle+\alpha_2\langle\sin^2\theta\rangle \nonumber \\
& =\alpha_2+(\alpha_1-\alpha_2)\langle\cos^2\theta\rangle,
\end{align}
where $\langle\cos^2\theta\rangle$ and $\langle\sin^2\theta\rangle$ are the expectation values of $\cos^2\theta$ and $\sin^2\theta$, respectively. 

On the other hand, when the applied electric field is parallel to the $y$-axis in Fig.\,\ref{V_molecule}(a), that is to say, the electric field is perpendicular to the alignment direction, the average molecular polarizability $\langle\alpha\rangle_\perp$ is given by
\begin{align}\label{eq_alpha_per}
\langle\alpha\rangle_{\perp,\text{3D}} & =\alpha_1\langle\cos^2\gamma\rangle+\alpha_2\langle\sin^2\gamma\rangle \nonumber \\
& =\alpha_2+(\alpha_1-\alpha_2)\langle\cos^2\gamma\rangle,
\end{align}
where $\gamma$ is the angle between the electric field, which is parallel to the $y$-axis in Fig.\,\ref{V_molecule}(a), and the long axis of the spheroidal molecule. Here, $\cos\gamma$ can be written as
\begin{equation}\label{eq_cosine_gamma}
\cos\gamma=\sin\theta\sin\phi,
\end{equation}
where $\phi$ is the angle between the $x$-axis and the direction of $\alpha_1$ projected onto the $xy$-plane. 

Now, $\langle\cos^2\theta\rangle_0$, which is the expectation value of $\cos^2\theta$ when the molecules are randomly oriented, is given by
\begin{equation}\label{eq_cosine_ave1}
\langle\cos^2\theta\rangle_0=\frac{\int_{0}^{\pi}\cos^2\theta d\Omega}{\int_{0}^{\pi} d\Omega},
\end{equation}
where $d\Omega$ is an infinitesimal solid angle, which is expressed as $2\pi \sin\theta d\theta d\phi$. Hence, by substituting $d\Omega = 2\pi \sin\theta d\theta d\phi$ into Eq.\,(\ref{eq_cosine_ave1}), we obtain 
\begin{equation}\label{eq_cosine_ave_3d}
\langle\cos^2\theta\rangle_{0,\text{3D}}=\frac{\int_{-\pi}^{\pi}\int_{0}^{\pi}2\pi\cos^2\theta\sin\theta d\theta d\phi}{\int_{-\pi}^{\pi}\int_{0}^{\pi}2\pi\sin\theta d\theta d\phi}=\frac{1}{3}.
\end{equation}
Similarly, $\langle\cos^2\gamma\rangle_0$, which is the expectation value of $\cos^2\gamma$ when the molecules are randomly oriented, is given by
\begin{align}\label{eq_cosine_gamma_ave_3d}
\langle\cos^2\gamma\rangle_{0} &=\frac{\int_{-\pi}^{\pi}\int_{0}^{\pi}2\pi\cos^2\gamma\sin\theta d\theta d\phi}{\int_{-\pi}^{\pi}\int_{0}^{\pi}2\pi\sin\theta d\theta d\phi}=\frac{1}{3}.
\end{align}
The mean polarizability of randomly oriented spheroidal molecules can thus be obtained, through substitution of Eq.\,(\ref{eq_cosine_ave_3d}) into Eq.\,(\ref{eq_alpha_par}) or substitution of Eq.\,(\ref{eq_cosine_gamma_ave_3d}) into Eq.\,(\ref{eq_alpha_per}), as
\begin{equation}\label{eq_alpha_3d}
\langle\alpha\rangle_{0,\text{3D}}=\frac{1}{3}\alpha_1+\frac{2}{3}\alpha_2.
\end{equation}

When the system is uniaxial, the distribution depends only on $\theta$. Since $\langle\cos^2\gamma\rangle$ does not depend on $\phi$ in this case, $\langle\cos^2\gamma\rangle$ is expressed as
\begin{align}\label{eq_cosine_gamma_ave_phi}
\langle\cos^2\gamma\rangle &=\frac{1}{2}\left( 1-\langle\cos^2\theta\rangle\right) .
\end{align}
As a result, Eq.\,(\ref{eq_alpha_per}) becomes
\begin{align}\label{eq_alpha_per_uniaxial}
\langle\alpha\rangle_{\perp,\text{3D}} & =\frac{1}{2}\left( \alpha_1+\alpha_2-\left( \alpha_1-\alpha_2\right) \langle\cos^2\theta\rangle\right) .
\end{align}
Therefore, the average polarizability for an ensemble of randomly orientated molecules in Eq.\,(\ref{eq_alpha_3d}) can be expressed in terms of $\langle\alpha\rangle_\parallel$ and $\langle\alpha\rangle_\perp$ as
\begin{equation}\label{eq_alpha_3d_par_per}
\langle\alpha\rangle_{0,\text{3D}}=\frac{1}{3}\langle\alpha\rangle_{\parallel,\text{3D}}+\frac{2}{3}\langle\alpha\rangle_{\perp,\text{3D}}.
\end{equation}

Here, we introduce the nematic order parameter, $S$, as a normalized degree of alignment. Namely, we require that $S =1$ for a perfectly aligned ensemble and $S =0$ for a randomly oriented ensemble.  $S$ can be expressed as an average of the long axis distribution of the angle $\theta$, which is the angle between a nanotube and the macroscopic alignment direction. For a 3D system~\cite{deGennesProst95Book},
\begin{align}\label{eq_order_para_3D}
S_{\text{3D}}&=\frac{1}{2}(3\langle\cos^2\theta\rangle-1)
\end{align}
%
satisfies the requirements above.  By reversing Eq.\,(\ref{eq_order_para_3D}), we obtain.
\begin{align}\label{eq_order_para_3D_cos}
\langle\cos^2\theta\rangle&=\frac{1}{3}(2S_{\text{3D}}+1).
\end{align}
The average polarizabilities for parallel and perpendicular electric fields, i.e., Eq.\,(\ref{eq_alpha_par}) and Eq.\,(\ref{eq_alpha_per}), can then be written in terms of $S$:
\begin{align}\label{eq_alpha_par_cos}
\langle\alpha\rangle_{\parallel,\text{3D}} =\frac{1}{3}\left \{ \alpha_1+2\alpha_2+2S_\text{3D}\left( \alpha_1-\alpha_2\right) \right \}.
\end{align}
\begin{align}\label{eq_alpha_per_cos}
\langle\alpha\rangle_{\perp,\text{3D}} =\frac{1}{3}\left \{ \alpha_1+2\alpha_2-S_\text{3D}\left( \alpha_1-\alpha_2\right) \right \}.
\end{align}

Given the average molecular polarizability, we can now obtain the susceptibility $\chi$ of the molecular ensemble as
\begin{equation}\label{susceptibility}
\chi = N\langle\alpha\rangle,
\end{equation}
where $N$ is the number of molecules. The absorption coefficient $\alpha_\text{abs}$ for incident light with angular frequency $\omega$ is then obtained by
\begin{align}\label{eq_absorption_coefficient}
\alpha_\text{abs} & = \frac{\omega}{cn_0}\chi'' = \frac{E_\text{ph}}{\hbar c n_0}\chi'' \nonumber \\
& = \frac{NE_\text{ph}}{\hbar c n_0}\langle\alpha''\rangle,
\end{align}
where $\chi''$ is the imaginary part of $\chi$, $E_\text{ph} = \hbar \omega$ is the photon energy of the incident light, $c$ is the speed of light, $\hbar$ is the reduced Planck constant, $n_0$ is the refractive index, and $\alpha''$ is the imaginary part of the molecular polarizability, $\alpha$. When the molecules are randomly oriented, $\alpha_\text{abs}$ can be obtained by substituting Eq.\,(\ref{eq_alpha_3d}) into Eq.\,(\ref{eq_absorption_coefficient}), i.e.,
\begin{align}\label{eq_absorption_coefficent_0_3D}
\alpha_{\text{abs},0,\text{3D}} = \frac{NE_\text{ph}}{3\hbar c n_0} \left( \alpha_1'' + 2\alpha_2'' \right),
\end{align}
where $\alpha_1''$ ($\alpha_2''$) is the imaginary part of $\alpha_1$ ($\alpha_2$).
Using Eq.\,(\ref{eq_alpha_par_cos}) and Eq.\,(\ref{eq_alpha_per_cos}), $\alpha_{\text{abs},\parallel,\text{3D}}$ and $\alpha_{\text{abs},\perp,\text{3D}}$, which are the absorption coefficients for parallel polarization and perpendicular polarization, respectively, can then be written as
\begin{align}\label{eq_absorptoin_par_cos}
\alpha_{\text{abs},\parallel,\text{3D}} =\frac{NE_\text{ph}}{3\hbar c n_0}\left\{ \alpha_1''+2\alpha_2''+2S_{\text{3D}}\left( \alpha_1''-\alpha_2''\right ) \right \}.
\end{align}
\begin{align}\label{eq_absorptoin_per_cos}
\alpha_{\text{abs},\perp,\text{3D}} =\frac{NE_\text{ph}}{3\hbar c n_0}\left \{ \alpha_1''+2\alpha_2''-S_{\text{3D}}\left( \alpha_1''-\alpha_2''\right) \right \}.
\end{align}
respectively. From Eqs.\,(\ref{eq_absorption_coefficent_0_3D}), (\ref{eq_absorptoin_par_cos}), and (\ref{eq_absorptoin_per_cos}), the following relation can also be derived:
\begin{align}\label{eq_absorption_coefficent_0_3D_par_per}
\alpha_{\text{abs},0,\text{3D}} =  \frac{1}{3}\alpha_{\text{abs},\parallel,\text{3D}}+\frac{2}{3}\alpha_{\text{abs},\perp,\text{3D}}.
\end{align}

When the reflection loss is negligible, the absorbance is given as $\alpha_\text{abs}l/\ln(10)$, where $l$ is the sample thickness. Therefore, the linear dichroism $LD$ is written as
\begin{align}\label{eq_LD_3D}
LD_{\text{3D}} &= \frac{l}{\ln(10)} (\alpha_{\text{abs},\parallel,\text{3D}} - \alpha_{\text{abs},\perp,\text{3D}})\\ &= \frac{NlE_\text{ph}}{\hbar c n_0 \ln(10)}S_{\text{3D}}\left( \alpha_1''-\alpha_2''\right).
\end{align}
The reduced linear dichroism $LD^r$, which is the linear dichroism normalized by $\alpha_{\text{abs},0,\text{3D}}l/\ln(10)$, where $\alpha_{\text{abs},0,\text{3D}}$ is given by Eq.\,(\ref{eq_absorption_coefficent_0_3D}) or Eq.\,(\ref{eq_absorption_coefficent_0_3D_par_per}). Thus,
\begin{align}\label{eq_LDr_3D}
LD_{\text{3D}}^r &= \frac{3\left(\alpha_{\text{abs},\parallel,\text{3D}} - \alpha_{\text{abs},\perp,\text{3D}}  \right) }{\alpha_{\text{abs},\parallel,\text{3D}}+2\alpha_{\text{abs},\perp,\text{3D}}}.
\end{align}
Substituting Eq.\,(\ref{eq_absorptoin_par_cos}) and Eq.\,(\ref{eq_absorptoin_per_cos}) here, we obtain
\begin{align}\label{eq_LDr_3D2}
LD_{\text{3D}}^r &= \frac{3S_{3\text{D}}(\alpha_1''-\alpha_2'')}{\alpha_1''+2\alpha_2''}.
\end{align}
Defining an angle $\xi_\text{3D} \equiv \tan^{-1} \left ( \sqrt{\alpha_1''/2\alpha_2''} \right )$,
\begin{align}\label{eq_LDr_3D3}
LD_{\text{3D}}^r &= \frac{1}{2}S\left( 3\cos^2\xi_\text{3D}-1\right) .
\end{align}

\subsection{Two-dimensional (2D) case}

We apply the above-developed 3D theory to an ensemble of planar or 2D aligned nanotubes.  As shown in Fig.\,\ref{V_molecule}(c), we define the polarizability along the tube axis as $\alpha_1$ and the polarizability perpendicular to the tube axis as $\alpha_2$. As before, $\theta$ is the angle between the macroscopic alignment direction and the individual nanotube under question. 

The expectation value of the polarizability of this 2D ensemble $\langle\alpha\rangle_\text{2D}$ for an electric field parallel to the alignment direction is given by
\begin{align}\label{eq_alpha_par_2D}
\langle\alpha\rangle_{\parallel,\text{2D}} & =\alpha_1\langle\cos^2\theta\rangle+\alpha_2\langle\sin^2\theta\rangle \nonumber \\
& =\alpha_2+(\alpha_1-\alpha_2)\langle\cos^2\theta\rangle,
\end{align}
and that for an electric field perpendicular to the alignment direction is given by
\begin{align}\label{eq_alpha_per_2D}
\langle\alpha\rangle_{\perp,\text{2D}} & =\alpha_1\langle\sin^2\theta\rangle+\alpha_2\langle\cos^2\theta\rangle \nonumber \\
& =\alpha_1+(\alpha_2-\alpha_1)\langle\cos^2\theta\rangle.
\end{align}

When the nanotubes are randomly oriented, the expectation value of $\cos^2\theta$ is given by
\begin{equation}\label{eq_cosine_ave_2D}
\langle\cos^2\theta\rangle_{0,\text{2D}}=\frac{\int_{0}^{\pi}\cos^2\theta d\theta}{\int_{0}^{\pi} d\theta}=\frac{1}{2}.
\end{equation}
The mean polarizability of randomly oriented nanotubes can then be obtained by substituting Eq.\,(\ref{eq_cosine_ave_2D}) into Eq.\,(\ref{eq_alpha_par_2D}) or Eq.\,(\ref{eq_alpha_per_2D}) as
\begin{equation}\label{eq_alpha_2d}
\langle\alpha\rangle_{0,\text{2D}}=\frac{1}{2}\alpha_1+\frac{1}{2}\alpha_2.
\end{equation}

The order parameter $S$ in 2D is expressed as \cite{Straley71PRA,FrenkelEppenga85PRA,Zamora-LedezmaetAl08NL},
\begin{equation}\label{eq_order_para_2D}
S_{\text{2D}}=\langle2\cos^2\theta-1\rangle.
\end{equation}
By reversing this equation, we obtain
\begin{equation}
\langle\cos^2\theta\rangle=\frac{1}{2}\left( S_{\text{2D}}+1\right) .
\end{equation}
The average polarizabilities for parallel and perpendicular electric fields, obtained as Eq.\,(\ref{eq_alpha_par_2D}) and Eq.\,(\ref{eq_alpha_per_2D}), respectively, can then be expressed in terms of $S_\text{2D}$ as
\begin{align}\label{eq_alpha_par_cos_2D}
\langle\alpha\rangle_{\parallel,\text{2D}} =\frac{1}{2}\left \{ \alpha_1+\alpha_2+S_{\text{2D}}\left( \alpha_1-\alpha_2\right) \right \}
\end{align}
and
\begin{align}\label{eq_alpha_per_cos_2D}
\langle\alpha\rangle_{\perp,\text{2D}} =\frac{1}{2}\left \{ \alpha_1+\alpha_2-S_{\text{2D}}\left( \alpha_1-\alpha_2\right) \right \}.
\end{align}
respectively.  

When the nanotubes are randomly oriented, the absorption coefficient $\alpha_\text{abs}$ can be obtained, by substituting Eq.\,(\ref{eq_alpha_2d}) into Eq.\,(\ref{eq_absorption_coefficient}), as
\begin{align}\label{eq_absorption_coefficent_0_2D}
\alpha_{\text{abs},0,\text{2D}} =  \frac{NE_\text{ph}}{2\hbar c n_0}\left(\alpha_1''+\alpha_2''\right),
\end{align}
where $\alpha_1''$ ($\alpha_2''$) is the imaginary part of $\alpha_1$ ($\alpha_2$).
From Eq.\,(\ref{eq_alpha_par_cos_2D}) and Eq.\,(\ref{eq_alpha_per_cos_2D}), the absorption coefficients for parallel and perpendicular polarizations are given, respectively, by
\begin{align}\label{eq_absorptoin_par_cos_2D}
\alpha_{\text{abs},\parallel,\text{2D}} =\frac{NE_\text{ph}}{2\hbar c n_0}\left \{ \alpha_1''+\alpha_2''+S_{\text{2D}}\left( \alpha_1''-\alpha_2''\right) \right \},
\end{align}
and
\begin{align}\label{eq_absorption_per_cos_2D}
\alpha_{\text{abs},\perp,\text{2D}} =\frac{NE_\text{ph}}{2\hbar c n_0}\left \{ \alpha_1''+\alpha_2''-S_{\text{2D}}\left( \alpha_1''-\alpha_2''\right )\right \}.
\end{align}
The absorption coefficient for randomly orientated nanotubes is also expressed by 
\begin{align}\label{eq_absorption_coefficent_0_2D_par_per}
\alpha_{\text{abs},0,\text{2D}} =  \frac{1}{2}\alpha_{\text{abs},\parallel,\text{2D}} + \frac{1}{2}\alpha_{\text{abs},\perp,\text{2D}}.
\end{align}
and
\begin{align}\label{eq_absorption_ratio_2D}
\frac{\alpha_{\text{abs},\parallel,\text{2D}}}{\alpha_{\text{abs},\perp,\text{2D}}} =\frac{\alpha_1''+\alpha_2''+S_{\text{2D}}\left( \alpha_1''-\alpha_2''\right)}{\alpha_1''+\alpha_2''-S_{\text{2D}}\left( \alpha_1''-\alpha_2''\right)}.
\end{align}

In a manner similar to the 3D case, the linear dichroism, $LD$, is expressed as
\begin{align}\label{eq_LD_2D}
LD_{\text{2D}} & = \frac{l}{\ln(10)} (\alpha_{\text{abs},\parallel,\text{2D}} - \alpha_{\text{abs},\perp,\text{2D}}) \\ &= \frac{NlE_\text{ph}}{2\hbar c n_0 \ln(10)}S_{\text{2D}}\left(\alpha_1''-\alpha_2''\right).
\end{align}
The reduced linear dichroism $LD^r$ is given by
\begin{align}\label{eq_LDr_2D}
LD_{\text{2D}}^r &= \frac{2 (\alpha_{\text{abs},\parallel,\text{2D}} - \alpha_{\text{abs},\perp,\text{2D}} ) }{\alpha_{\text{abs},\parallel,\text{2D}}+\alpha_{\text{abs},\perp,\text{2D}}} .
\end{align}
Substituting Eq.\,(\ref{eq_absorptoin_par_cos_2D}) and Eq.\,(\ref{eq_absorption_per_cos_2D}) here, we obtain
\begin{align}\label{eq_LDr_2D2}
LD_{\text{2D}}^r &= \frac{2S_{\text{2D}}(\alpha_1''-\alpha_2'')}{\alpha_1''+\alpha_2''}.
\end{align}
Defining an angle $\xi_\text{2D} \equiv \tan^{-1} \left ( \sqrt{\alpha_1''/\alpha_2''} \right )$,
\begin{align}\label{eq_LDr_2D3}
LD_{\text{2D}}^r &= 2S_\text{2D}\left( \cos^2\xi_\text{2D}-1\right) .
\end{align}

Finally, we consider absorption coefficients for two cases: (i)~$\xi_\text{2D}=0^\circ$ ($\alpha_1'' \ne 0$, $\alpha_2'' = 0$), and (ii)~$\xi_\text{2D}=90^\circ$ ($\alpha_1'' = 0$, $\alpha_2'' \ne 0$). In these cases, $\alpha_{\text{abs},\parallel}$, $\alpha_{\text{abs},\perp}$, $\alpha_{\text{abs},\parallel}/\alpha_{\text{abs},\perp}$, and $LD^r$ are expressed as follows:\\

\noindent (i)~$\xi_\text{2D}=0^\circ$ ($\alpha_1'' \ne 0$, $\alpha_2'' = 0$)
\begin{align}
\alpha_{\text{abs},\parallel,\text{2D}} = \frac{NE_\text{ph}\alpha_1''}{2\hbar c n_0}\left( 1+S_{\text{2D}} \right)\\
\label{eq_absorption_par_alpha1}
\alpha_{\text{abs},\perp,\text{2D}} = \frac{NE_\text{ph}\alpha_1''}{2\hbar c n_0}\left( 1-S_{\text{2D}} \right)\\
\frac{\alpha_{\text{abs},\parallel,\text{2D}}}{\alpha_{\text{abs},\perp,\text{2D}}} = \frac{1+S_{\text{2D}}}{1-S_{\text{2D}}}\\
LD_{\text{2D}}^r = 2S_{\text{2D}}.
\end{align}
%
(ii)~$\xi_\text{2D}=90^\circ$ ($\alpha_1'' = 0$, $\alpha_2'' \ne 0$)
\begin{align}\label{eq_absorption_par_alpha2}
\alpha_{\text{abs},\parallel,\text{2D}} =\frac{NE_\text{ph}\alpha_2''}{2\hbar c n_0}\left( 1-S_{\text{2D}} \right)\\
\alpha_{\text{abs},\perp,\text{2D}} =\frac{NE_\text{ph}\alpha_2''}{2\hbar c n_0}\left( 1+S_{\text{2D}} \right)\\
\frac{\alpha_{\text{abs},\parallel,\text{2D}}}{\alpha_{\text{abs},\perp,\text{2D}}} =\frac{1-S_{\text{2D}}}{1+S_{\text{2D}}}\\
LD_{\text{2D}}^r =-2S_{\text{2D}}.
\end{align}


\end{document}